\documentclass[a4paper]{article}
\usepackage{indentfirst}
\usepackage{fullpage}
\usepackage{amsmath}
\usepackage{amsthm}
\usepackage{latexsym}
\usepackage{epsf}
\usepackage{graphicx,color}

\makeatletter
\renewcommand\paragraph{%
  \@startsection{paragraph}{4}{0mm}%
   {-\baselineskip}%
   {.5\baselineskip}%
   {\normalfont\normalsize\bfseries}}
\makeatother

\begin{document}
% 
% \theoremstyle{remark}
% \newtheorem{remark}{Remark}[section]
% \numberwithin{equation}{section}
\newtheorem{lemma}{Lemma}[section]

\newcommand{\bA}{{\bf A}}
\newcommand{\B}{{\bf B}}
\def\bC{{\bf C}}
\newcommand{\bco}{{\boldsymbol{:}}}
\newcommand{\blambda}{{\boldsymbol{\lambda}}}
\newcommand{\bLambda}{{\boldsymbol{\Lambda}}}
\newcommand{\bmu}{{\boldsymbol{\mu}}}
\newcommand{\bn}{{\bf n}}
\newcommand{\bnabla}{{\boldsymbol{\nabla}}}
\newcommand{\bomega}{{\boldsymbol{\omega}}}
\newcommand{\bsigma}{{\boldsymbol{\sigma}}}
\newcommand{\btheta}{{\boldsymbol{\theta}}}
\newcommand{\bo}{{\overline{B(0)}}}
\newcommand{\bu}{{\bf u}}
\newcommand{\bU}{{\bf U}}
\newcommand{\bv}{{\bf v}}
\newcommand{\bw}{{\bf w}}
\newcommand{\bzero}{{\bf 0}}
\newcommand{\ct}{{\mathcal{T}}}
\newcommand{\cth}{{\mathcal{T}_h}}
\newcommand{\dsum}{{\displaystyle\sum}}
\newcommand{\bD}{{\bf D}}
\newcommand{\e}{{\bf e}}
\newcommand{\bF}{{\bf F}}
\newcommand{\bbf}{{\bf f}}
\newcommand{\bG}{{\bf G}}
\newcommand{\bg}{{\bf g}}
\newcommand{\Gx}{{{\overrightarrow{\bf Gx}}}}
\def\hs2{{\hskip -2pt}}

\newcommand{\bI}{{\bf I}} 
\newcommand{\bL}{{\bf L}}
\newcommand{\intbt}{{\displaystyle{\int_{\small B(t)}}}}

\newcommand{\intG}{{\displaystyle{\int_{\Gamma}}}}
\newcommand{\into}{{\displaystyle{\int_{\Omega}}}}
\newcommand{\intobt}{{\displaystyle{\int_{\Omega\setminus\overline{B(t)}}}}}
\newcommand{\intpb}{{\displaystyle{\int_{\partial B}}}}
\newcommand{\lto}{{L^2(\Omega)}}
\newcommand{\no}{{\noindent}}
\newcommand{\obo}{{\Omega \backslash \overline{B(0)}}}
\newcommand{\obt}{{\Omega \backslash \overline{B(t)}}}
\newcommand{\oo}{{\overline{\Omega}}}
\newcommand{\R}{{\text{I\!R}}}
\newcommand{\bs}{{\bf s}}
\newcommand{\bT}{{\bf T}}
\newcommand{\bV}{{\bf V}}
\newcommand{\W}{{\bf W}}
\newcommand{\bx}{{\bf x}}
\newcommand{\bxi}{{\boldsymbol{\xi}}}
\newcommand{\bY}{{\bf Y}}
\newcommand{\bby}{{\bf y}}
\newcommand{\bz}{{\bf z}}
\newcommand{\br}{{\bar{y}}}
\newcommand{\bbu}{{\bar{u}}}

\title{\Large \bf \boldmath\  Dynamics of particle sedimentation in viscoelastic fluids: 
A numerical study on particle chain in two-dimensional narrow channel}

\author{Tsorng-Whay Pan\footnotemark[1]\\
{\it Department of Mathematics, University of Houston, Houston, Texas 77204, USA} \\ \\
  Roland Glowinski\\
{\it Department of Mathematics, University of Houston, Houston, Texas 77204, USA}\\
{\it Department of Mathematics, Hong Kong Baptist University, Hong Kong}}

\renewcommand{\thefootnote}{\fnsymbol{footnote}}
%\footnotetext[1]{lingling@math.uh.edu}
\footnotetext[1]{pan@math.uh.edu}
%\footnotetext[3]{roland@math.uh.edu}
%\footnotetext[2]{Project supported by an NSF grant DMS-9973318.}
\date{}
\maketitle

\begin{abstract}
In this article we present a numerical method for simulating the sedimentation of circular particles
in two-dimensional channel filled with a viscoelastic fluid of FENE-CR type, which is generalized
from a domain/distributed Lagrange multiplier method with a factorization approach for Oldroyd-B fluids 
developed in [J. Non-Newtonian Fluid Mech. 156 (2009) 95].   Numerical results suggest that the polymer extension 
limit $L$ for the FENE-CR fluid has no effect on the final formation of vertical chain for the cases 
of two disks and three disks in two-dimensional narrow channel, at least for the values of $L$ considered 
in this article; but the intermediate dynamics  of particle interaction before having a vertical chain 
can be different for the smaller values of $L$ when increasing the relaxation time. For the cases of six particles sedimenting 
in FENE-CR type viscoelastic fluid, the formation of  chain of 4 to 6 disks does depend on the polymer extension limit $L$. 
For the smaller values of $L$, FENE-CR type viscoelastic fluid can not bring them together like the case of these 
particles settling in a vertical chain formation in Oldroyd-B fluid; but  two separated chains of three disks are formed.  
{Similar results for the case of ten disks are also obtained. The numerical results of several more particle cases 
suggest  that for smaller values of $L$, the length of the vertical chain is shorter and the size of cluster is smaller.} 

\vskip 1ex


\end{abstract}

\baselineskip 14pt

\setlength{\parindent}{1.5em}

\section{Introduction}

\normalsize
 
The motion of particles in non-Newtonian fluids is not only of fundamental theoretical interest,
but is also of importance in many applications to industrial processes involving particle-laden 
materials (see, e.g., \cite{chhabra1993} and \cite{mckinley2002}).
For example,  during the hydraulic fracturing operation used in oil and gas wells, 
suspensions of solid particles in polymeric solutions are pumped into hydraulically-induced 
fractures. The particles must prop these channels open to enhance the rate of oil
recovery \cite{Economides1989}. During the shut-in stage, proppant settling is pronounced
when the fluid pressure decreases due to the end of hydraulic fracturing process.  The 
study of particle chain during settling in vertical channel can help us to understand 
the mechanism of proppant agglomeration in narrow  fracture zones \cite{tomac2015}.

Although numerical methods for simulating particulate flows in Newtonian fluids have been 
very successful, numerically simulating  particulate flows in viscoelastic
fluids is much more complicated and challenging. One of the difficulties (e.g., see 
\cite{baaijens1998}, \cite{keunings2000}) for simulating viscoelastic flows is the breakdown
of the numerical methods. It has been widely believed that the lack of positive definiteness
preserving property of the conformation tensor at the discrete level during the {\it entire
time integration} is one of the reasons for the breakdown. To preserve the positive definiteness
property of the conformation tensor,  several methodologies  have been proposed recently, as 
in \cite{fattal2004}, \cite{fattal2005}, \cite{lee2006} and \cite{lozinski2003}.
Lozinski and Owens \cite{lozinski2003} factored the conformation
tensor to get $\bsigma =A A^T$ and then they wrote down the equations for $A$ approximately at 
the discrete level. Hence, the positive definiteness of the conformation tensor is forced with 
such an approach. The methodologies developed in   \cite{lozinski2003} have been applied 
in \cite{Hao2009} together with the FD/DLM method through operator splitting techniques for 
simulating particulate flows in Oldroyd-B fluid. We have generalized these 
computational methodologies to viscoelastic fluids of the FENE-CR type, which is a more ``realistic'' 
model when compared  with the Oldroyd-B model as advocated in \cite{Hinch1988}. 
To study the effect of the polymer extension limit $L$ on the particle  chain formation while settling,
we have considered the cases of two, three and six disks settling in viscoelastic fluid is previously considered in \cite{Hao2009}
since vertical chains are known to be formed for these cases.  In this article, we have compared
the particle   sedimenting in a vertical two-dimensional channel filled with viscoelastic fluid of either Oldroyd-B 
or  FENE-CR types to find out the effect of the extension limit of the immersed polymer coils on the chaining.  
The computational results of disks settling in Oldroyd-B fluid are obtained by the numerical method developed in \cite{Hao2009}.
For the cases of two disks and three disks in two-dimensional narrow channel, numerical results suggest that the polymer extension 
limit $L$ for the FENE-CR fluid has no effect on the final formation of vertical particle chain l, at least for the values of $L$ considered 
in this article; but the intermediate particle dynamics can be different for the smaller values of $L$ when 
increasing the value of the relaxation time.  For six particles sedimenting in FENE-CR type  viscoelastic fluid, the formation of 
chain of 4 to 6 disks does depend on the polymer extension limit $L$. For the smaller values of $L$, FENE-CR 
type viscoelastic fluid can not bring them together like the case of these  particles settling in a vertical chain 
formation in Oldroyd-B fluid; but instead two separated chains of three disks are formed.
{Similar results for the case of ten disks are also obtained. The numerical results of several more particle cases suggest 
that for smaller values of $L$, the length of the vertical chain is shorter and the size of cluster is smaller.}
The article is organized as follows. In Section 2, we present a FD/DLM formulation for particulate flows in an FENE-CR fluid and 
the associated  the operator splitting technique, the space and time discretization of the FD/DLM formulation, how we
apply the Lozinski and Owens’ method to get the equivalent equations for the conformation tensor. In Section 3, numerical 
results for the cases of sedimentation of  two, three, and six particles and their chaining under the effect of the the polymer extension
are discussed.

\section{Mathematical Formulations and numerical methods}\label{sec2}

\subsection{Governing equations and its FD/DLM Formulation}
\begin{figure}[!tp]
\begin{center}
\leavevmode
\epsfxsize=2.5in
\epsffile{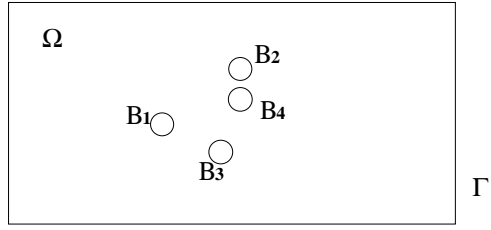}\\
\caption{An example of a two-dimensional flow region with four circular particles.} \label{fig1a}
\end{center}
\end{figure}

Following the work developed  in \cite{Hao2009},  we will first address in the following 
the models and computational methodologies combined with the Lozinski and Owens' factorization approach.
Let $\Omega$ be a bounded two-dimensional (2D) domain and  let $\Gamma$ be the boundary of $\Omega$.
We suppose that $\Omega$ is filled with a viscoelastic fluid of either Oldroyd-B
or FENE-CR type of density $\rho_f$ and that it contains $N$ moving rigid particles of density
$\rho_s$ (see Figure \ref{fig1a}). 
Let $B(t)=\displaystyle\cup_{i=1}^N B_i(t)$ where $B_i(t)$   is the $i$th
rigid particle in the fluid for $i=1,\dots,N$. We denote by $\partial B_i(t)$  the 
boundary  of  $B_i(t)$ for $i=1,\dots,N$.
For some $T>0,$ the governing equations for the fluid-particle system   are
\begin{eqnarray}
&& \hskip -20pt \rho_{f}( \dfrac{\partial \bu }{\partial t} +
( \bu \cdot \bnabla )\bu ) 
=\rho_{f}\bg -\bnabla p +2\mu \bnabla \cdot \bD(\bu)
+\bnabla \cdot \bsigma^p  \ \  in \ \Omega \backslash \overline{B(t)},\, t\in(0,T), \label{eqn:2.1.1}\\
&& \hskip -20pt \ \nabla \cdot \textbf{u}=0 \ \  in \ \Omega \backslash \overline{B(t)},\, t\in(0,T),\label{eqn:2.1.2}\\
&& \hskip -20pt \ \bu(\bx,0)=\bu_{0}(\bx), \ \  \forall \bx \in \Omega \backslash \overline{B(0)}, \ 
with \, \nabla \cdot \bu_{0}=0, \label{eqn:2.1.3}\\
&& \hskip -20pt \ \bu=\bg_{0} \ \  on \ \Gamma \times
(0, T), with \int _{\Gamma} \bg_{0} \cdot \bn\,
d \Gamma = 0,\label{eqn:2.1.4}\\
&& \hskip -20pt \ \bu= {\bV}_{p,i} + \omega_i \times \stackrel{\longrightarrow}{\bG_i\bx},
 \ \forall \bx \in  \partial B_i(t), \, i=1,\cdots,N,\label{eqn:2.1.5}\\
% \end{eqnarray}
% \begin{eqnarray}
&& \hskip -20pt  \dfrac{\partial \bC }{\partial t}+(\bu \cdot \bnabla)\ \bC
-(\bnabla \bu)\ \bC-\bC \ (\bnabla \bu)^t=-\dfrac{f(\bC)}{\lambda_{1}}(\bC -\bI) \,\,\, in \ \Omega \backslash \overline{B(t)}, \, t\in(0,T),\label{eqn:2.1.6}\\
&& \hskip -20pt \ \bC(\bx,0)=\bC_{0}(\bx),\,  \bx \in \Omega \backslash \overline{B(0)},\label{eqn:2.1.7}\\
&& \hskip -20pt \ \bC=\bC_{L}, \ \ on \ \Gamma^{-},\label{eqn:2.1.8}
\end{eqnarray}
where $\bu$ is the flow velocity, $p$ is the pressure,
$\bg$ is the gravity,   $\mu=\eta_1\lambda_2/\lambda_1$
is the solvent viscosity of the fluid, $\eta=\eta_1-\mu$ is the
elastic viscosity of the fluid, $\eta_1$ is the fluid viscosity,
$\lambda_1$ is the relaxation time of the fluid, $\lambda_2$ is the retardation
time of the fluid, $\bn$ is the outer normal unit vector at $\Gamma,$
$\Gamma^{-}$ is the upstream portion of $\Gamma$.
The  polymeric stress tensor $\bsigma^p$ in 
(\ref{eqn:2.1.1}) is given by $\bsigma^p=\dfrac{\eta}{\lambda_1} f(\bC) (\bC-\bI)$, where 
the conformation tensor $\bC$ is symmetric and positive definite (see \cite{joseph1990}) 
and $\bI$ is the identity matrix.
Setting $f$ equal to unity corresponds to the Oldroyd-B model while 
\begin{equation}
f(\bC)=\dfrac{L^2}{L^2-tr(\bC)} \label{eqn:2.1.32}
\end{equation}
corresponds to the FENE-CR model \cite{Chilcott1988}, where $tr(\bC)$ is the trace of 
the conformation tensor $\bC$  and  $L$ is the maximum extension of the immersed polymer coils
and referred to as the extensibility of the immersed polymer coils. The Oldroyd-B model then is a 
special case associated with infinite extensibility.

In (\ref{eqn:2.1.5}), the no-slip condition holds on the boundary  of the $i$th particle,  
$\bV_{p,i}$ is the translation velocity,   $\omega_i$ is the angular velocity and
$\bG_i$ is the center of mass and  $\omega_i \times \stackrel{\longrightarrow}{\bG_i\bx}=(-\omega_i (x_2 - G_{i,2}), \omega_i (x_1 - G_{i,1}))$
(for 2D cases  considered in this article).
The motion of the  particles is modeled by Newton's laws:
\begin{eqnarray} 
&& M_{p,i} \ \dfrac{d \bV_{p,i}}{d t} = M_{p,i}\bg + \bF_i + \bF_i^r,\label{eqn:2.1.9} \\
&&  I_{p,i} \dfrac{d  \omega_i }{d t}= F^t_i,\label{eqn:2.1.10}\\
&& \dfrac{d \bG_i}{d t} = \bV_{p,i},\label{eqn:2.1.11}\\
&& \bG_i(0) = \bG^0_i, \bV_{p,i}(0) = \bV^0_{p,i}, \omega_i(0) = {\omega}^0_i,\label{eqn:2.1.12}
\end{eqnarray} 
for $i=1, \dots, N$,  where in (\ref{eqn:2.1.9})-(\ref{eqn:2.1.12}),  
$M_{p,i}$ and $I_{p,i}$    are the  the mass and the   inertia of the $i$th particle, 
respectively, $\bF^r_i$ is a short range repulsion  force imposed on the $i$th particle
by other particles and the wall to prevent particle/particle and particle/wall penetration
(see \cite{glowinski2001} for details), and
$\bF_i $ and $ {F}^t_i$ denote the hydrodynamic force and the associated torque
imposed on the $i$th particle by the fluid, respectively.

To avoid the frequent remeshing and the difficulty of the  mesh generation for a time-varying domain 
in which the rigid particles can be very close to each other, especially for 
three dimensional particulate flow, we have extended the governing equations 
to the entire domain $\Omega$ (a fictitious domain). For a fictitious-domain-based variational formulation of 
the governing equations of the particulate flow,  we consider 
only one rigid particle $B(t)$ (a disk in 2D) in the fluid domain without losing generality. 
Let us  define first the following functional spaces
\begin{eqnarray}
&& \bV_{\bg_0(t)}=\{ \ \bv\ |\ \bv\ \in ({H^1(\Omega)})^2, \bv=\bg_0(t) \ on\ \Gamma
\}, \nonumber \\
&& L_0^2(\Omega)=\{ \ q\ |\ q\ \in {L^2(\Omega)}, \int_{\Omega}
q\, d\bx=0 \}, \nonumber \\
&& \bV_{\bC_{L}(t)}=\{ \ \bC\ |\ \bC\ \in ({H^1(\Omega)})^{2 \times 2},
 \bC=\bC_{L}(t) \ on\ \Gamma^{-}\}, \nonumber \\
&& \bV_{\bC_{0}}=\{ \ \bC\ |\ \bC\ \in ({H^1(\Omega)})^{2 \times 2},
 \bC=0 \ on\ \Gamma^{-}\}, \nonumber \\
&& \Lambda(t)={H^1(B(t))}^2. \nonumber
\end{eqnarray}
Following the methodologies developed in \cite{glowinski2001,singh2000}, a fictitious domain 
formulation of the governing equations  (\ref{eqn:2.1.1})-(\ref{eqn:2.1.12}) reads as follows:

{\em For a.e. $t>0,$ find  $\bu(t) \in \bV_{\bg_0(t)}$, $p(t)\in  L_0^2(\Omega)$, $\bC(t)\in
\bV_{\bC_{L}(t)}$, $\bV(t)\in \R^2$, $\bG(t)\in \R^2$, $\omega(t) \in \R$,
 $\blambda(t) \in \Lambda(t)$  such that}
\begin{eqnarray}
&& \begin{cases}
\displaystyle \rho_{f}\int_{\Omega} \left[ \dfrac{\partial \bu}{\partial t} +
(\bu \cdot \bnabla)\bu \right] \cdot \bv\, d\bx 
+2\mu \int_{\Omega}\bD (\bu):\bD (\bv)\,\, d\bx - \int_{\Omega} p \bnabla \cdot \bv\, d\bx \\
\displaystyle -\int_{\Omega} \bv \cdot  (\bnabla \cdot \bsigma^p )\,\, d\bx +
 (1-\rho_f/\rho_s) \lbrace M_p \dfrac{d\bV}{dt} \cdot \bY 
 + I_p \dfrac{d\omega}{dt}\cdot \theta \rbrace \\
\displaystyle -<\blambda, \bv-\bY -\theta \times {\stackrel{\longrightarrow}{\bG\bx}} >_{B(t)} 
 -\bF^r \cdot \bY \\
\displaystyle =\rho_{f} \int_{\Omega} \mathbf{g} \cdot \mathbf{v}d \mathbf{x} 
+(1-\rho_f/\rho_s)M_p \bg \cdot \bY,\\
\displaystyle \forall \{\bv, \bY, \theta\} \in 
(H_0^1(\Omega))^2 \times \R^2 \times \R,
\end{cases}  \label{eqn:2.1.24}\\
% \end{eqnarray}
% \begin{eqnarray}
&& \int_{\Omega} q \bnabla \cdot \bu(t) \, d \bx = 0, 
 \forall  q \in L^2(\Omega),  \label{eqn:2.1.25} \\ 
&& <\bmu, \bu(\bx, t) - \bV(t) - \omega(t) \times {\stackrel{\longrightarrow}{\bG(t)\bx}} >_{B(t)}=0,  \ \ \forall \bmu  \in  \bLambda(t),  \label{eqn:2.1.26}\\
&&\int_{\Omega} \left( \dfrac{\partial \bC }{\partial t}+(\bu \cdot \bnabla)\bC 
-(\bnabla \bu)\bC-\bC (\bnabla \bu)^t \right) 
:\bs \, d\bx \label{eqn:2.1.27}\\
&& = -\int_{\Omega} \dfrac{f(\bC)}{\lambda_{1}}(\bC- \bI):\bs \, d\bx,
\forall \bs \in \bV_{\bC_0}, \ with\ \ \bC=\bI \ \ in\ \ B(t), \nonumber \\
&& \dfrac{d \bG}{dt} =\bV,   \label{eqn:2.1.28} \\
&& \bC(\bx, 0) = \bC_0(\bx), \forall \bx \in \Omega,  \ with\ \ \bC_0=\bI \ \ in\ \ B(0),  \label{eqn:2.1.29}\\ 
&& \bG(0) = \bG_0, \ \bV(0) = \bV_0, \ \omega(0) = \omega_0, \ B(0) = B_0, \label{eqn:2.1.30}\\
&& \bu(\bx, 0) = \begin{cases}\bu_0(\bx), \forall \bx \in \Omega\setminus{\overline {B_0}}, \\
 \bV_0 + \omega_0 \times {\stackrel{\longrightarrow}{\bG_0\bx}}, \ \forall \bx \in {\overline{B_0}}.
\end{cases} \label{eqn:2.1.31}
\end{eqnarray}
In (\ref{eqn:2.1.24}) the {\it Lagrange multiplier} $\blambda$  defined over $B$ can be viewed as
an extra body force maintaining the rigid body motion inside $B$. The conformation tensor
$\bC$ inside the rigid particle is extended as the identity tensor $\bI$
as in (\ref{eqn:2.1.27}) since the polymeric stress tensor is zero inside the rigid particle.
In equation (\ref{eqn:2.1.24}), since $\bu$ is divergence free and satisfies the Dirichlet 
boundary conditions on $\Gamma,$ we have 
$2 \int_{\Omega}\bD (\bu):\bD (\bv)
\, d\bx =\int_{\Omega}  \bnabla \bu: \bnabla \bv \, d\bx, \, \forall \bv \in (H_0^1(\Omega))^2$.
This is a substantial simplification from the computational point of view, which is another
advantage of the fictitious domain approach. With this simplification, we can use, as shown in the following 
section, fast solvers for the elliptic problems in order to speed up computations.
Also the gravity term $\bg$ in ($\ref{eqn:2.1.24}$) can be absorbed in the pressure term. 

\subsection{Finite Element Approximation}
In order to solve problem (\ref {eqn:2.1.24})-(\ref {eqn:2.1.31}) numerically, we shall discretize
the fictitious domain $\Omega$ using an uniform finite element mesh $\ct_h$ for the
velocity and conformation tensor, where $h$ is the mesh size, and a 
twice coarser uniform mesh $\ct_{2h}$ for the pressure. 
The following finite dimensional spaces are defined for approximating $\bV_{\bg_0(t)}$, 
$ (H_0^1(\Omega))^2$, $L^2(\Omega)$, $L_0^2(\Omega)$,
 $\bV_{\bC_{L}(t)}$, $\bV_{\bC_0}$, respectively,
\begin{eqnarray*}
&& \bV_{\bg_{0h}(t)}=\{ \ \bv_h\ |\ \bv_h\ \in ({C^0(\overline{\Omega})})^2, 
\bv_h |_E\in (P_1)^2, \forall E \in \ct_h,  \bv_h|_\Gamma=\bg_{0h}(t) 
\},  \\
&& \bV_{0h}=\{ \ \bv_h\ |\ \bv_h\ \in ({C^0(\overline{\Omega})})^2, 
\bv_h |_E\in (P_1)^2, \forall E \in \ct_h,  \bv_h|_\Gamma=0 
\},  \\
&& L_h^2=\{ \ q_h\ |\ q_h\ \in C^0(\overline{\Omega}), 
q_h |_E\in P_1, \forall E \in \ct_{2h}\},  \\
&& L_{0h}^2=\{ \ q_h\ |\ q_h\ \in {L^2_h}, \int_{\Omega}
q_h\ \, d\bx={\bf 0} \},  \\
% \end{eqnarray*}
% \begin{eqnarray*}
&& \bV_{\bC_{Lh}(t)}=\{ \ s_h\ |\ s_h\ \in ({C^0(\overline{\Omega})})^{2\times 2}, 
s_h |_E\in (P_1)^{2\times 2}, \forall E \in \ct_h,  s_h|_{\Gamma_h^{-}}=\bC_{Lh}(t) \},  \\
&& \bV_{\bC_{0h}}=\{ \ s_h\ |\ s_h\ \in ({C^0(\overline{\Omega})})^{2\times 2}, 
s_h |_E\in (P_1)^{2\times 2}, \forall E \in \ct_h,  s_h|_{\Gamma_h^{-}}=0 \}
\end{eqnarray*}
where $P_1$ is the space of the polynomials in two variables of degree $\le1,$
 $\bg_{0h}(t)$ is an approximation of $\bg_0$ satisfying $ \int_{\Gamma}
\bg_{0h}(t)\cdot \bn d\Gamma=0,$ and $\Gamma_h^{-}=\{\bx \ |\ \bx 
\in \Gamma,\ \bg_{0h}(\bx,t)\cdot \bn(\bx)<0\}. $
The discrete Lagrange multiplier space $ \Lambda_h(t)$ is defined as follows: 
let 
${\{\bx_i\}}^{K}_{i=1}$ be a set of points from $\overline{B(t)}$ 
that covers $\overline{B(t)}$ evenly, and then we define
\begin{equation} 
\bLambda_h(t)= \{ \bmu \ | \ \bmu = \sum_{j=1}^K \bmu_j \delta(\bx - \bx_j), 
    \ \bmu_j \in \R^2, \ \forall j= 1,..., K\},  \label{eqn:3.1.1}
\end{equation}  
where $\bx \to \delta(\bx - \bx_j)$ is the Dirac measure at $\bx_j$.
Then instead of the scalar product of ${(H^1(B_h(t))}^2 $, we shall
use $ < \bmu , \bv >_{B_h(t)}$ defined by
\begin{equation} 
< \bmu , \bv >_{B_h(t)} = \sum_{j=1}^K \bmu_j \cdot \bv(\bx_j), 
\forall \bmu \in \bLambda_h(t), \, \bv \in \bV_{\bg_{0h}(t)} \, or\, \bV_{0h}. 
\label{eqn:3.1.2}
\end{equation}  
Using the above scalar product implies that
the rigid body motion of $B(t)$ is forced via a {\it collocation method} \cite{glowinski2001}. 
% The  set $\{\bx_j\}_{j=1}^K$ is the union of two subsets, namely: (i) The set of the points of 
% the velocity grid contained in $B(t)$ and whose distance at $\partial B(t)$ is no less than $ch$, $h$ 
% being a space discretization step and  $c$ a constant $\approx 1$. (ii) A set of control points 
% located on $\partial B(t)$ and forming a mesh whose step size is of the order of $h$.

Then a discrete analogue of the problem (\ref {eqn:2.1.24})-(\ref {eqn:2.1.31}) can be 
obtained with the above finite dimensional spaces.
\kern -0.3in
\subsection{ An Operator Splitting Scheme}
\kern -0.1in
Consider the following initial value problem:
\begin{equation}
 \dfrac{d \phi}{d t} + A(\phi) = 0 \ {\rm on} \ (0, T), \ \
\phi(0)=\phi_0 \label{eqn:3.1.11}
\end{equation}
with $0 < T < +\infty$. We suppose that operator $A$ has a decomposition such as
$A = \sum_{j=1}^J A_j$ with $J \ge 2$. 
Let $\tau (> 0)$ be a time-discretization step,  
 we denote $n\tau$ by $t^n$. With $\phi^n$ denoting an approximation of 
 $\phi(t^n)$, the {\bf Lie scheme} \cite{chorin1978} reads as follows:
 
% \begin{equation} 
% \phi^0=\phi_0;   \label{eqn:3.1.12}
% \end{equation} 
For $n \ge 0$, assuming that $\phi^n$  is known (with $\phi^0=\phi_0$), compute $\phi^{n+1}$ via 
\begin{equation}
\begin{cases}
&\dfrac{d\phi}{dt}  + A_j(\phi) = 0 \ \ on \ (t^n, t^{n+1}),   \\ 
&\phi(t^n) = \phi^{n+(j-1)/J}; \phi^{n+j/J} =\phi(t^{n+1}),  
\end{cases}\label{eqn:3.1.13}  
\end{equation}
for $j = 1, \dots, J$. The Lie's scheme is first order
accurate, but its low order of accuracy is compensated by its simplicity, making
it (relatively) easy to implement, and by its robustness. Some classical operator 
splitting techniques with application to the Navier-Stokes equations have been 
discussed in \cite{glowinski2003} in details.

The Lie's operator splitting scheme allows us to decouple the following
difficulties:\\
(1). The incompressibility condition, and the related unknown pressure;\\
(2). The advection terms;\\
(3). The rigid-body motion in $B_h(t)$, and the related DLM $\blambda_h$.

The constitutive equation satisfied by the conformation tensor $\bC$ is split with $J=3$ for
now to show how the factorization approach works out. Suppose that  $\bC^n$ and $\bu$ are 
known, we compute
\begin{eqnarray}
&&\begin{cases}
\dfrac{d\bC}{dt}  + (\bu \cdot \bnabla)\bC = 0 \ \ on \  \ (t^n, t^{n+1}),\\
\bC(t^n)=\bC^n; \ \bC^{n+1/3}=\bC(t^{n+1}),
\end{cases}\label{eqn:3.1.13a}  \\
&&\begin{cases}
\dfrac{d\bC}{dt} - (\bnabla \bu)\bC-\bC(\bnabla \bu)^t + \dfrac{f(\bC^{n+1/3})}{\lambda_1}\bC = 0
 \ \ on \  \ (t^n, t^{n+1}), \\
 \bC(t^n)=\bC^{n+1/3};\ \bC^{n+2/3}=\bC(t^{n+1}),
 \end{cases}\label{eqn:3.1.13b}  \\
&& \begin{cases}
\dfrac{d\bC}{dt} = \dfrac{f(\bC^{n+2/3})}{\lambda_1}\bI \ \ on \  \ (t^n, t^{n+1}), \\
\bC(t^n)=\bC^{n+2/3};\ \bC^{n+1}=\bC(t^{n+1}).
 \end{cases}\label{eqn:3.1.13c}  
\end{eqnarray}
We have derived the following two equivalent equations based on the factorization approach
with $s=f(\bC^{n+1/3})$ for equations (\ref {eqn:3.1.13a}) and  (\ref {eqn:3.1.13b}):
\begin{lemma}
 For a matrix $A$ and $\bC=A A^t$, given the velocity $\bu$, $\lambda_1 (>0)$ and a constant $s$,\\
(a). if $A$ satisfies the equation $\dfrac{dA}{dt}  + (\bu \cdot \bnabla)A = 0$, 
then $\bC$ satisfies  the equation
 $$\dfrac{d\bC}{dt}  + (\bu \cdot \bnabla)\bC = 0;$$ 
(b). if $A$ satisfies the equation
 $\dfrac{dA}{dt}+ \dfrac{s}{2\lambda_1}A  -( \bnabla \bu)A = 0$, 
then $\bC$ satisfies  the equation
 $$\dfrac{d\bC}{dt} + \dfrac{s}{\lambda_1}\bC 
- (\bnabla \bu)\bC-\bC(\bnabla \bu)^t = 0.$$
\end{lemma}

\noindent {\it Proof:} (a) Multiplying the equation by $A^t$ to the right, and 
the transpose of the equation by $A$ to the left,  we have,
\begin{displaymath} 
\dfrac{dA}{dt}A^t  + (\bu \cdot \bnabla)A A^t= 0, \,\,\,\,\,(L1)
\end{displaymath}
\begin{displaymath} 
A\dfrac{dA^t}{dt}  + A(\bu \cdot \bnabla)A^t= 0,\,\,\,\,\, (L2)
\end{displaymath}
Adding (L1) and (L2) gives,
\begin{displaymath} 
\dfrac{d(AA^t)}{dt}  + (\bu \cdot \bnabla)(AA^t)= 0; \text{  that is},
 \dfrac{d\bC}{dt}  + (\bu \cdot \bnabla)(\bC)= 0.
\end{displaymath}
(b) Multiplying the equation by $A^t$ to the right, and 
the transpose of the equation by $A$ to the left,  we have,
\begin{displaymath} 
\dfrac{dA}{dt}A^t  +\dfrac{s}{2\lambda_1}AA^t -(\bnabla \bu)AA^t= 0, \,\,\,\,\,(L3)
\end{displaymath}
\begin{displaymath} 
A\dfrac{dA^t}{dt}  +\dfrac{s}{2\lambda_1}AA^t - AA^t{(\bnabla \bu)}^t= 0,\,\,\,\,\, (L4)
\end{displaymath}
Adding (L3) and (L4) gives,
\begin{displaymath} 
\dfrac{d(AA^t)}{dt}  +\dfrac{s}{\lambda_1}AA^t -(\bnabla \bu)AA^t-AA^t{(\bnabla \bu)}^t= 0,
\end{displaymath}
or,
\begin{displaymath} 
\dfrac{d \bC}{dt}  +\dfrac{s}{\lambda_1}\bC -(\bnabla \bu)\bC-\bC{(\bnabla \bu)}^t= 0. \hfill \qed
\end{displaymath}

Similarly, we can define finite 
dimensional spaces $\bV_{A_{Lh}(t)}$ and $\bV_{A_{0h}}$ for $A$. 
When applying  the Lie's scheme to the discrete analogue of the problem (\ref {eqn:2.1.24})-(\ref {eqn:2.1.31})
with the above factorization
and equations for $A$ and the backward Euler's method to some sub-problems, we obtain
\begin{equation} 
\bu^0={\bu}_{0h},\bC^{0}={\bC}_{0h},
 \bG^0=\bG_0, \bV^0=\bV_0, \omega^0=\omega_0 \,\, \text{given},\label{eqn:3.1.14} 
\end{equation}
for $n \ge 0$, $\bu^n, \bC^{n}, \bG^n, \bV^n, \omega^n \,$ being known,
we compute $\bu^{n+\frac{1}{5}}$, and $p^{n+\frac{1}{5}}$ via the solution of  
\begin{equation}
\begin{cases}
\displaystyle \rho_{f}\int_{\Omega}  \dfrac{ \bu^{n+\frac{1}{5}}-\bu^n}
{\triangle t} \cdot \bv \, d\bx
 - \int_{\Omega} p^{n+\frac{1}{5}} \bnabla \cdot \bv \, d\bx =0,
\forall \bv \in \bV_{0h}  \\
\displaystyle \int_{\Omega} q \bnabla \cdot \bu^{n+\frac{1}{5}} \, d \bx = 0, 
 \forall  q \in L^2_h; \bu^{n+\frac{1}{5}} \in \bV^{n+1}_{\bg_{0h}},
 p^{n+\frac{1}{5}} \in L^2_{0h}.
\end{cases} \label{eqn:3.1.15}
\end{equation}
Next, we compute $\bu^{n+\frac{2}{5}}$ and $A^{n+\frac{2}{5}}$ via the solution of
\begin{equation}
\begin{cases}
\displaystyle \rho_{f}\int_{\Omega}  \dfrac{d\bu(t)}{dt} \cdot \bv\, d \bx 
+ \int_{\Omega} (\bu^{n+\frac{1}{5}} \cdot \bnabla)\bu(t) \cdot \bv \, d\bx =0,
 \forall \bv \in \bV_{0h}^{n+1,-};\\
\displaystyle \bu(t^n)=\bu^{n+\frac{1}{5}},\\
\displaystyle \bu(t) \in \bV_h, \bu(t)=\bg_{0h}(t^{n+1})\ on \ 
\Gamma^{n+1,-} \times[t^n,t^{n+1}] ;
\end{cases} \label{eqn:3.1.16}
\end{equation}
\begin{equation}
\begin{cases}
\displaystyle \int_{\Omega}  \dfrac{dA(t)}{dt} : \bs\, d \bx + \int_{\Omega} (\bu^{n+\frac{1}{5}} \cdot \bnabla)
 A(t):\bs \, d\bx =0, \forall \bs \in \bV_{A_{0h}};\\
\displaystyle A(t^n)=A^n, \,\, where \, A^n{(A^n)}^t=\bC^{n}\\
\displaystyle A(t) \in \bV^{n+1}_{A_{Lh}},\, t \in [t^n,t^{n+1}] ;
\end{cases} \label{eqn:3.1.17}
\end{equation}
and set $\bu^{n+\frac{2}{5}}=\bu(t^{n+1})$ and $A^{n+\frac{2}{5}}=A(t^{n+1})$,  
 where $\Gamma^{n+1,-}=\lbrace \bx \in \Gamma,\bg_{0h}(t^{n+1})(\bx) \cdot \bn(\bx) < 0
\rbrace,$ $\bV_{h}=\{ \bv_h|\bv_h \in (C^0(\overline{\Omega}))^2, 
\bv_h |_E\in (P_1)^2, \forall E \in \ct_h,\}$, and $\bV_{0h}^{n+1,-}=\lbrace \bv \in \bV_h, \bv=0, \, on \, 
\Gamma^{n+1,-} \rbrace.$

Then, compute $\bu^{n+\frac{3}{5}}$ and $A^{n+\frac{3}{5}}$ via the solution of
\begin{equation}
\begin{cases}
\displaystyle \rho_{f}\int_{\Omega}  
\dfrac{\bu^{n+\frac{3}{5}}-\bu^{n+\frac{2}{5}}}{\triangle t} \cdot \bv\, d \bx +\alpha \mu \int_{\Omega} \bnabla \bu^{n+\frac{3}{5}}
 : \bnabla \bv \, d\bx =0,\\
\displaystyle \forall \bv \in \bV_{0h}; \bu^{n+\frac{3}{5}} \in \bV_{\bg_{0h}}^{n+1},
\end{cases} \label{eqn:3.1.18}
\end{equation}

\begin{equation}
\begin{cases}
\displaystyle \int_{\Omega}(
 \dfrac{A^{n+\frac{3}{5}}-A^{n+\frac{2}{5}}}{\triangle t} 
-(\bnabla \bu^{n+\frac{3}{5}})A^{n+\frac{3}{5}}
+\dfrac{f(\bA^{n+\frac{2}{5}}(\bA^{n+\frac{2}{5}})^t)}{2\lambda_1}A^{n+\frac{3}{5}}): \bs \, d\bx =0,\\
\displaystyle \forall \bs \in \bV_{A_{0h}};
A^{n+\frac{3}{5}} \in \bV^{n+1}_{A_{Lh}},
\end{cases} \label{eqn:3.1.19}
\end{equation}
and set 
\begin{equation}
\bC^{n+\frac{3}{5}}=A^{n+\frac{3}{5}}(A^{n+\frac{3}{5}})^t
+\dfrac{ \triangle t \ f(\bA^{n+\frac{3}{5}}(\bA^{n+\frac{3}{5}})^t)}
{\lambda_1}\bI.\label{eqn:3.1.19a}
\end{equation}
Then, predict the position and the translation velocity of the center 
of mass as follows:

   Take $\bV^{n+\frac{3}{5},0}=\bV^n$ and $\bG^{n+\frac{3}{5},0}=\bG^n$;
then predict the new position and translation velocity via the
following sub-cycling and predicting-correcting technique

\noindent For $k=1,2,\dots,N,$ compute
 \begin{eqnarray} 
&& {\hat{\bV}}^{n+\frac{3}{5},k}=\bV^{n+\frac{3}{5},k-1}  +(1-\rho_f/\rho_s)^{-1}M_p^{-1}\bF^r(\bG^{n+\frac{3}{5},k-1})
\triangle t/2N, \label{eqn:3.1.20} \\
&&{\hat \bG}^{n+\frac{3}{5},k}=\bG^{n+\frac{3}{5},k-1}+ (\triangle t/4N)
({\hat{\bV}}^{n+\frac{3}{5},k}+\bV^{n+\frac{3}{5},k-1}), \label{eqn:3.1.21}\\
&& \bV^{n+\frac{3}{5},k}=\bV^{n+\frac{3}{5},k-1}  +(1-\rho_f/\rho_s)^{-1}M_p^{-1}(\bF^r({\hat \bG}^{n+\frac{3}{5},k})
+\bF^r(\bG^{n+\frac{3}{5},k-1}))\triangle t/4N,  \label{eqn:3.1.22}  \\
&& \bG^{n+\frac{3}{5},k}=\bG^{n+\frac{3}{5},k-1}+ (\triangle t/4N)
(V^{n+\frac{3}{5},k}+V^{n+\frac{3}{5},k-1}), \label{eqn:3.1.23}
\end{eqnarray} 
end do;\ \ let $\bV^{n+\frac{3}{5}}=\bV^{n+\frac{3}{5},N}$,
 $\bG^{n+\frac{3}{5}}=\bG^{n+\frac{3}{5},N}$.
 
\noindent Next compute $\{ \bu^{n+\frac{4}{5}},\blambda^{n+\frac{4}{5}},\bV^{n+\frac{4}{5}},
\omega^{n+\frac{4}{5}} \}$ via the solution of  
\begin{eqnarray}
&&\begin{cases}
\displaystyle \rho_{f}\int_{\Omega}
\dfrac{\bu^{n+\frac{4}{5}}-\bu^{n+\frac{3}{5}}}{\triangle t}
   \cdot \mathbf{v} d\mathbf{x} +\beta \mu \int_{\Omega}\bnabla 
\bu^{n+\frac{4}{5}} : \bnabla \bv \, d\bx \\
\displaystyle + 
 (1-\dfrac{\rho_f}{\rho_s}) \left[ M_p
\dfrac{\bV^{n+\frac{4}{5}}-\bV^{n+\frac{3}{5}}}{\triangle t} \cdot \bY 
 + I_p \dfrac{\omega^{n+\frac{4}{5}}-\omega^{n}}{\triangle t}
\cdot \theta \right] \\
\displaystyle 
= <\blambda^{n+\frac{4}{5}}, \bv-\bY -\theta \times \bG^{n+\frac{3}{5}}
 \bx>_{B_h^{n+\frac{3}{5}}} + (1-\rho_f/\rho_s) M_p \bg \cdot \bY, \\
\displaystyle \forall \bv \in \bV_{0h},\bY \in \R^2, \theta \in \R,\\
\ <\bmu, \bu^{n+\frac{4}{5}} - \bV^{n+\frac{4}{5}} - \omega^{n+\frac{4}{5}} 
\times \bG^{n+\frac{3}{5}}\bx>_{B_h^{n+\frac{3}{5}}}=0,   
 \forall \bmu  \in  \bLambda_h^{n+\frac{3}{5}};\\
\bu^{n+\frac{4}{5}} \in \bV_{\bg_{0h}}^{n+1},
\blambda^{n+\frac{4}{5}} \in \Lambda_h^{n+\frac{3}{5}},
\end{cases} \label{eqn:3.1.24}
\end{eqnarray}
and set $\bC^{n+\frac{4}{5}}=\bC^{n+\frac{3}{5}}$, and then let
$\bC^{n+\frac{4}{5}}=\bI$ in $B_h^{n+\frac{3}{5}}$.

Then take  $\bV^{n+1,0}=\bV^{n+\frac{4}{5}}$ and 
$\bG^{n+1,0}=\bG^{n+\frac{3}{5}}$; and predict the final position and
 translation velocity as follows:

\noindent For $k=1,2,\dots,N,$ compute
 \begin{eqnarray} 
&& {\hat{\bV}}^{n+1,k}=\bV^{n+1,k-1} 
+(1-\rho_f/\rho_s)^{-1}M^{-1}\bF^r(\bG^{n+1,k-1})
\triangle t/2N,\label{eqn:3.1.25}\\
&&{\hat \bG}^{n+1,k}=\bG^{n+1,k-1}+ (\triangle t/4N)
({\hat{\bV}}^{n+1,k}+\bV^{n+1,k-1}), \label{eqn:3.1.26}\\
&& \bV^{n+1,k}=\bV^{n+1,k-1} +(1-\rho_f/\rho_s)^{-1}M^{-1}(\bF^r({\hat \bG}^{n+1,k})
+\bF^r(\bG^{n+1,k-1}))\triangle t/4N,  \label{eqn:3.1.27} \\
&& \bG^{n+1,k}=\bG^{n+1,k-1}+ (\triangle t/4N)
(\bV^{n+1,k}+\bV^{n+1,k-1}), \label{eqn:3.1.28}
\end{eqnarray} 
end do;\ \ let $\bV^{n+1}=\bV^{n+1,N}$, $\bG^{n+1}=\bG^{n+1,N}$.

Finally, compute $\bu^{n+1}$ via the solution of
\begin{equation}
\begin{cases}
\displaystyle \rho_{f}\int_{\Omega}  
\dfrac{\bu^{n+1}-\bu^{n+\frac{4}{5}}}{\triangle t} \cdot \bv\, d \bx +\gamma \mu \int_{\Omega} \bnabla \bu^{n+1}
 : \bnabla \bv \, d\bx ,\\
\displaystyle =\dfrac{\eta}{\lambda_1}\int_{\Omega} \bv \cdot (\bnabla \cdot   f(\bC^{n+\frac{4}{5}})(\bC^{n+\frac{4}{5}}-\bI))
 \, d\bx,
 \forall \bv \in \bV_{0h}; \bu^{n+1} \in \bV_{\bg_{0h}}^{n+1}.
\end{cases} \label{eqn:3.1.29}
\end{equation}
We complete the final step by setting $\bC^{n+1}=\bC^{n+\frac{4}{5}}$,
and $\omega^{n+1}=\omega^{n+\frac{4}{5}}$. 

\noindent In the above, $\bu_{0h}$ is an approximation of $\bu_0$ so that 
$\int_{\Omega} q \bnabla \cdot \bu_{0h}d\bx=0,\ \forall q\in L^2_h$, 
$\bV_{\bg_{0h}}^{n+1}=\bV_{\bg_{0h}(t^{n+1})}$,
$\Lambda_h^{n+s}=\Lambda_h(t^{n+s})$,
$\bV_{A_{Lh}}^{n+1}=\bV_{A_{Lh}(t^{n+1})}$, 
$B_h^{n+s}=B_h(t^{n+s})$,  the spaces $\bV_{\bA_{L_h}(t)}$ and $\bV_{\bA_{0h}}$ for $\bA$ 
are defined similar to those $\bC_{L_h}(t)$ and $\bC_{0h}$, and 
$\alpha + \beta +\gamma=1,$ for $\alpha,\beta,\gamma \ge 0$.

\begin{figure} [!tp]
\begin{minipage}{4.2in}
%         \begin{figure}
\includegraphics[width=4.0in]{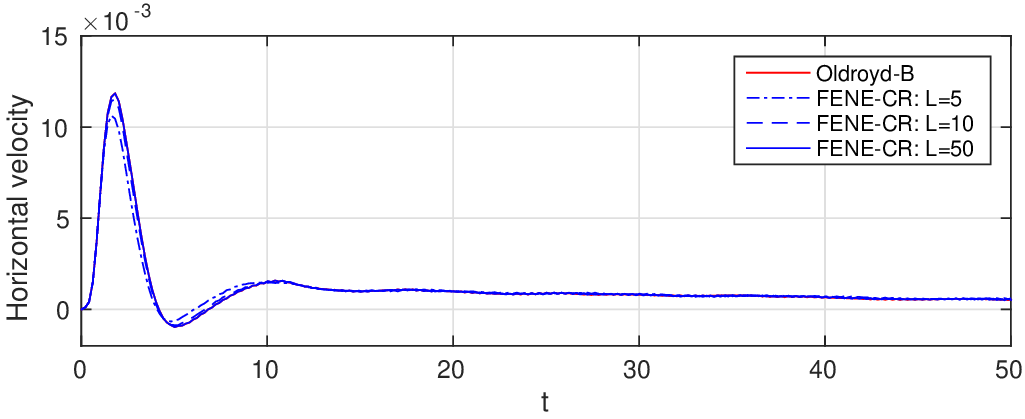}\\
\includegraphics[width=4.0in]{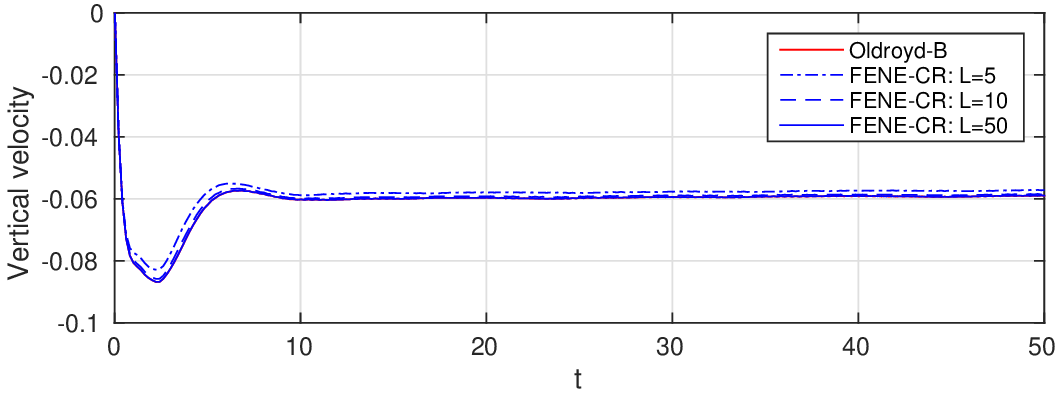}\
%         \end{figure}
  \end{minipage}      
  \begin{minipage}{2in}
%         \begin{figure}
\includegraphics[width=1.8in]{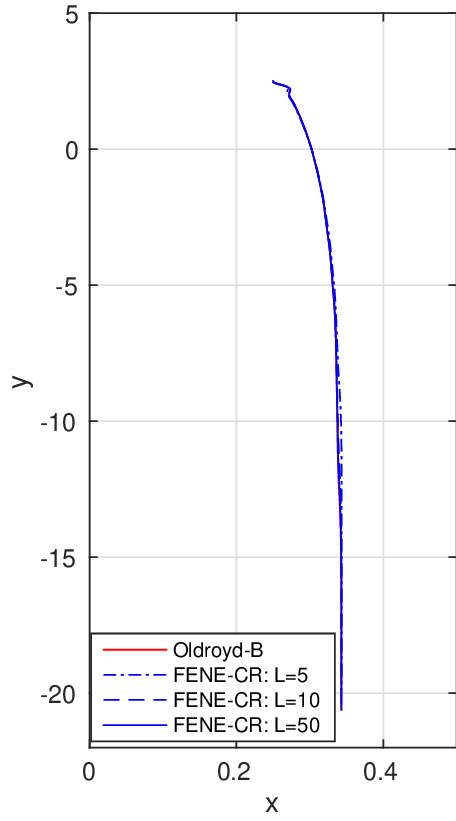}  
%         \end{figure}
  \end{minipage}       
\caption{Histories of the particle horizontal velocity (left top), vertical velocity (left bottom) and trajectories of a disk (right) for $\lambda_1$=2.025
(the associated numbers are Re=0.4186, M=0.4393, De=0.4611, E=1.1016 for Oldroyd-B fluid and  Re=0.4059, M=0.4261, De=0.4472, 
E=1.1016 for FENE-CR fluid of L=5). }  \label{fig1b}
% \end{figure}   
% \begin{figure}  
\begin{minipage}{4.2in}
%         \begin{figure}
\includegraphics[width=4.0in]{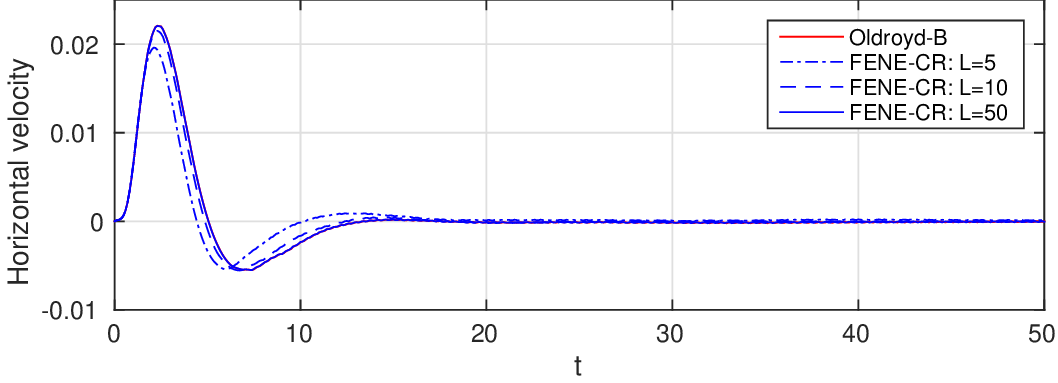}\\
\includegraphics[width=4.0in]{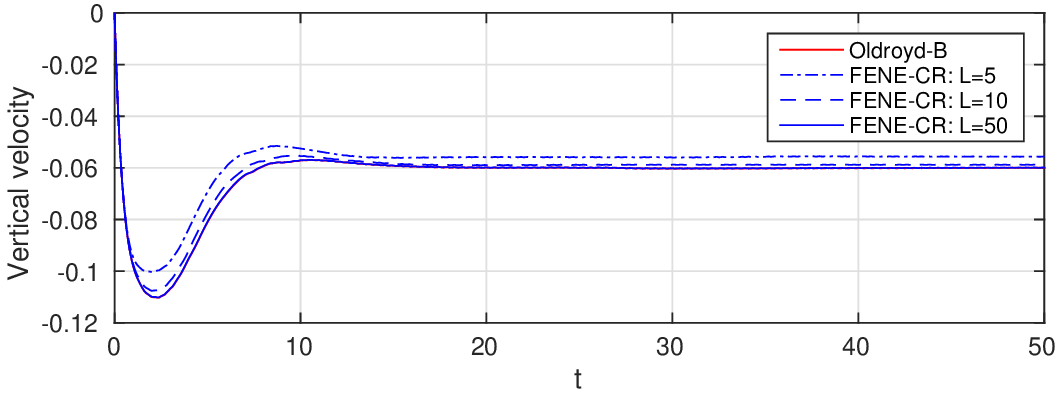}\
%         \end{figure}
  \end{minipage}      
  \begin{minipage}{2in}
%         \begin{figure}
\includegraphics[width=1.8in]{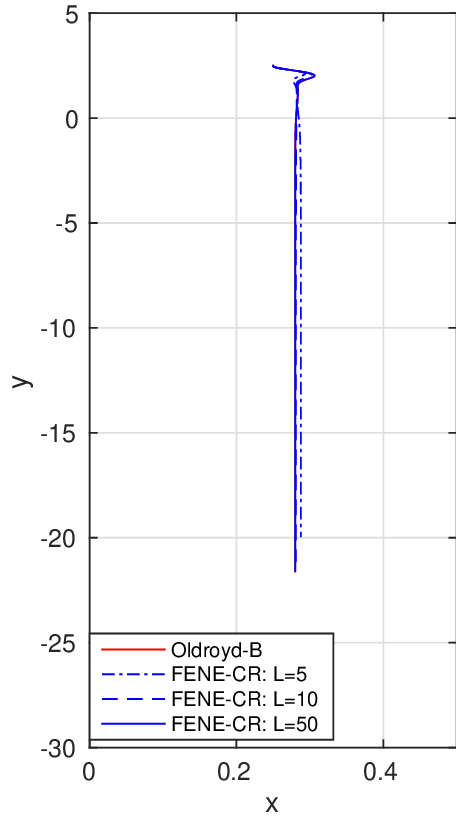}  
%         \end{figure}
  \end{minipage}       
\caption{Histories of the particle horizontal velocity (left top), vertical velocity (left bottom) and trajectories of a disk (right) for $\lambda_1$=4.05
(the associated numbers are Re=0.4403, M=0.6536, De=0.9701, E=2.2032 for Oldroyd-B fluid and  Re=0.4110, M=0.6101, De=0.9056, 
E=2.2032 for FENE-CR fluid of L=5). }  \label{fig1c}
\end{figure}   

\subsection{Solution strategies}

In the algorithm (\ref{eqn:3.1.14})-(\ref{eqn:3.1.29}), we have obtained a sequence of simpler sub-problems,
 namely:  
(i) using a $L^2$-projection Stokes solver \`a la  Uzawa to force the incompressibility condition  in (\ref{eqn:3.1.15}), 
(ii) an advection step  for the velocity and conformation tensor in (\ref{eqn:3.1.16}) and (\ref{eqn:3.1.17}),
(iii) a diffusion step  for the velocity in (\ref{eqn:3.1.18}) and the step for the rest of
the constitutive equations for  the conformation tensor   in (\ref{eqn:3.1.19}) and (\ref{eqn:3.1.19a}), 
(iv) a step to predict the particle position in (\ref{eqn:3.1.20})-(\ref{eqn:3.1.23}), 
(v) a step to enforce the rigid body motion inside the particle and to obtain its updated translation 
and angular velocity in (\ref{eqn:3.1.24}) and then to set the conformation tensor to be an identity matrix inside the particle,
(vi) a step to correct the particle position in (\ref{eqn:3.1.25})-(\ref{eqn:3.1.28}), 
and (vii) a diffusion step with the updated polymeric stress tensor for the velocity in (\ref{eqn:3.1.29}).

The resulting methodology is easy to implement and quite modular. 
A degenerated quasi-Stokes problem (\ref{eqn:3.1.15})  is solved by an Uzawa/ preconditioned 
conjugate gradient algorithm operating in the space $L^2_{0h}$ discussed in \cite{Hao2009} and \cite{glowinski2003}.
The advection problems (\ref{eqn:3.1.16}) and (\ref{eqn:3.1.17}) are solved by a wave-like equation method
(see  \cite{glowinski2003} and   \cite{dean1997})  which is an explicit method and does not introduce numerical 
dissipation. Since the advection problem is decoupled from the other ones, we can choose a proper sub-time step 
so that the CFL condition is satisfied.  Problem (\ref{eqn:3.1.24}), concerning the rigid body motion enforcement, is a 
saddle point problem and is solved by a conjugate gradient method given in  e.g., \cite{Hao2009} and \cite{glowinski2003}.
Problems (\ref{eqn:3.1.18}) and (\ref{eqn:3.1.29}) are classical elliptic problems which can be solved by a matrix-free fast solver.
In (\ref{eqn:3.1.20})-(\ref{eqn:3.1.23}) and  (\ref{eqn:3.1.25})-(\ref{eqn:3.1.28}), it is
a predicting-correcting scheme to obtain the position
of the mass center and the translation velocity of the particle.
Problem (\ref{eqn:3.1.19})  gives a simple equation at each grid point
which can be solved easily if we use the trapezoidal quadrature rule to compute the integrals as
in  \cite{Hao2009}.

\section{Numerical Results and discussion}

To study the effect of the polymer extension limit $L$ on the particle  chain formation while settling,
we have considered the cases of two, three and six disks settling in viscoelastic fluid as in \cite{Hao2009}
since vertical chains are known to be formed for these cases.  The computational results for disks settling in Oldroyd-B fluid 
are obtained by the numerical method developed in \cite{Hao2009} and these results are compared with those results obtained by
the scheme discussed in the previous section for the FENE-CR model at large values of $L$ for validation purpose since  
$f(\bC)=\dfrac{L^2}{L^2-tr(\bC)} \to 1$ as $L \to \infty$ 
(i.e., the  FENE-CR model has almost recovered the Oldroyd-B model for the large values of $L$). 
In the following discussion, the  particle Reynolds number is Re$=\dfrac{\rho_f U d}{\eta_1}$, the Debra number is De$=\dfrac{\lambda_1 U}{d}$, 
the Mack number is M=$\sqrt{\rm De Re}$, and the elasticity number is E=De/Re=$\dfrac{\lambda_1 \eta_1}{ d^2 \rho_f}$ where $U$ is 
the averaged terminal speed of disks and $d$ is the disk diameter.

\begin{figure}
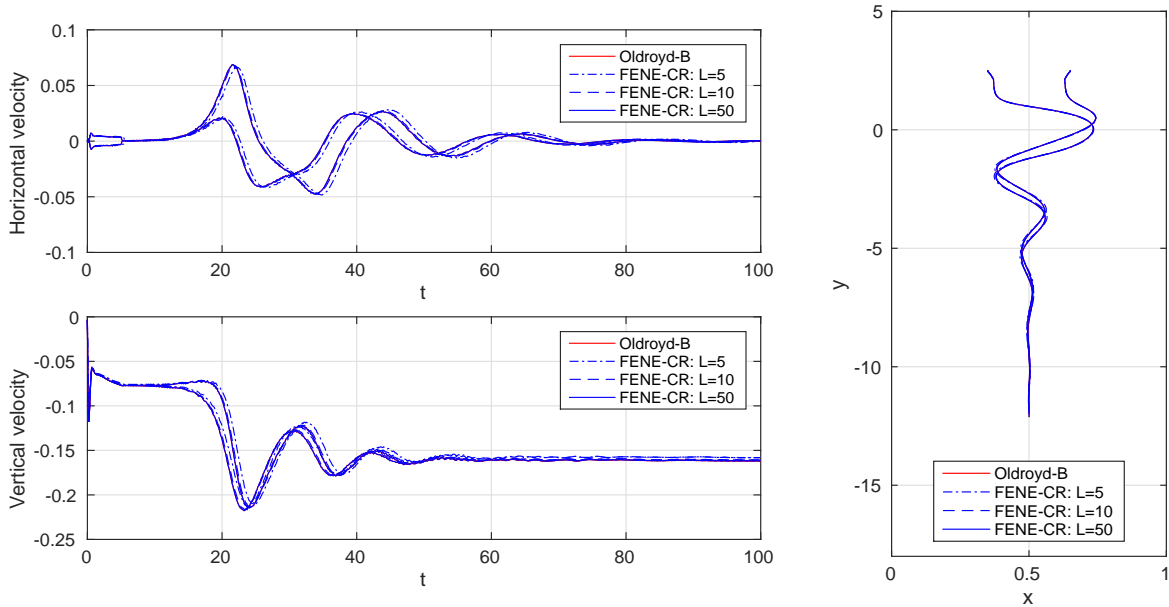
 
\begin{minipage}{4.2in}
%         \begin{figure}
\includegraphics[width=4.0in]{2disk-case1E1d6-xcrv-comp.eps}\\
\includegraphics[width=4.0in]{2disk-case1E1d6-ycrv-comp.eps}\
%         \end{figure}
  \end{minipage}      
  \begin{minipage}{2in}
%         \begin{figure}
\includegraphics[width=1.8in]{2disk-case1E1d6-position-comp.eps}  
%         \end{figure}
  \end{minipage}       
\caption{Histories of the particle horizontal velocity (left top), vertical velocity (left bottom) and trajectories of two disks (right) for $\lambda_1$=0.5
(the associated numbers are Re=0.202, M=0.255, De=0.323, E=1.6 and  Re=0.197, M=0.249, De=0.316, E=1.6 for Oldroyd-B fluid and FENE-CR 
of L=5, respectively). }  \label{fig2a}
\end{figure}

\begin{figure}
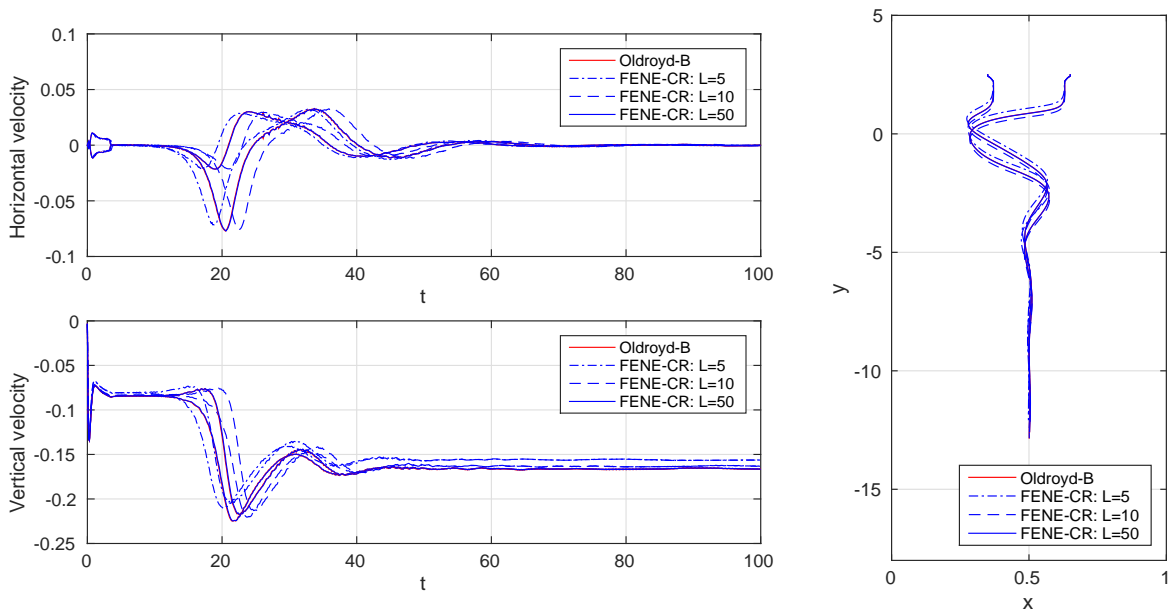
 
\begin{minipage}{4.2in}
%         \begin{figure}
\includegraphics[width=4.0in]{2disk-case1E3d2-xcrv-comp.eps}\\
\includegraphics[width=4.0in]{2disk-case1E3d2-ycrv-comp.eps}\
%         \end{figure}
  \end{minipage}      
  \begin{minipage}{2in}
%         \begin{figure}
\includegraphics[width=1.8in]{2disk-case1E3d2-position-comp.eps}  
%         \end{figure}
  \end{minipage}       
\caption{Histories of the particle horizontal velocity (left top), vertical velocity (left bottom) and trajectories of  two disks (right) for $\lambda_1$=1
(the associated numbers are Re=0.208, M=0.371, De=0.664, E=3.2 and Re=0.195, M=0.35, De=0.625, E=3.2 for Oldroyd-B fluid and FENE-CR 
of L=5, respectively).}  \label{fig2b}
\end{figure}   
  
\subsection{Few settling disks}

We have first considered the  settling of one disk in a vertical channel of  infinite length filled with a viscoelastic fluid as in \cite{Hao2009}, 
the computational domain is $\Omega= (0,1)\times(0,6)$ initially and then it moves vertically with the mass center of the 
disk  (see, e.g., \cite{Hu1992} and \cite{Pan2002} and references therein for adjusting the 
computational domain according to the position of the particle). The disk diameter is $d=$0.25 
and the initial position of the disk center is at (0.25, 2.5).  The disk density $\rho_s$ is 1.0007 and the fluid 
density $\rho_f$ is 1. The fluid viscosity $\eta_1$ is 0.034.  The relaxation time $\lambda_1$ is either 2.025 or 4.05 
and the retardation time $\lambda_2$ is $\lambda_1/8$. Hence the values of the elasticity number E are 1.1016 and 2.2032 for 
$\lambda_1=2.025$ and 4.05, respectively. The maximal polymer extension $L$ is either 5, 10, or 50 for the  FENE-CR model. 
The mesh sizes for the velocity field, conformation tensor and pressure are $h=1/128$, 1/128, and 1/64, respectively; and  
the time step is 0.0004. Fig. \ref{fig1b} shows that  the trajectories of a disk settling in either Oldroyd-B or FENE-CR 
fluids are almost identical  for  $L$=5, 10, and 50 and $\lambda_1=1.1016$ (E=1.1016); but in Fig. \ref{fig1c}  the  disk trajectory for 
$L$=5 and $\lambda_1=4.05$   is quite different from those for  $L=10$ and 50 and $\lambda_1=4.05$ and that 
for Oldroyd-B fluid. For $\lambda_1=4.05$ (E=2.2032), the  disk trajectory for $L$=5  is different from others is due to that 
its Deborah number is slightly smaller (actually the values of associated De are 0.9056, 0.9519, 0.9693, and 0.9701
for the FENE-CR fluid for $L$=5, 10, 50, and Oldroyd-B). As observed in \cite{Feng1996}, 
the effect of Deborah number on the single particle settling trajectory is that the smaller De is, the closer to the 
center line the equilibrium position is. Histories of the particle horizontal velocity, vertical velocity and trajectories of a disk obtained for this 
one disk case show that the larger value of  polymer extension limit $L$ is, the closer to the those of a disk settling in  Oldroyd-B fluid is.

\begin{figure}
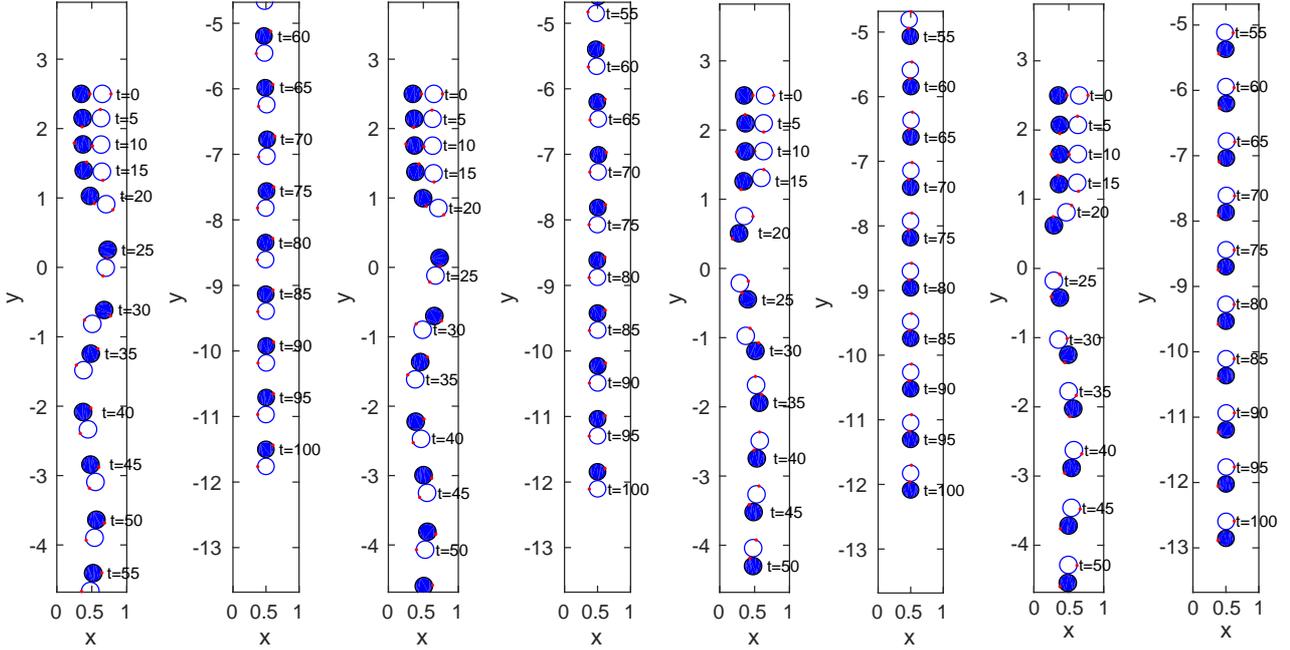

     \begin{center} 
     \hskip -20pt \includegraphics[width=0.125\textwidth]{2disk-case1E1d6-L005-1.eps}
     \includegraphics[width=0.125\textwidth]{2disk-case1E1d6-L005-2.eps}
     \includegraphics[width=0.125\textwidth]{2disk-case1E1d6-OL-1.eps}
     \includegraphics[width=0.125\textwidth]{2disk-case1E1d6-OL-2.eps}     
     \includegraphics[width=0.118\textwidth]{2disk-case1E3d2-L005-1.eps}
     \includegraphics[width=0.118\textwidth]{2disk-case1E3d2-L005-2.eps}
     \includegraphics[width=0.118\textwidth]{2disk-case1E3d2-OL-1.eps}
     \includegraphics[width=0.118\textwidth]{2disk-case1E3d2-OL-2.eps}         
     \end{center}  
     \caption{Positions of two disks :  FENE-CR of $L=$5 and $\lambda_1=0.5$ (left two),  Oldroyd-B and $\lambda_1=0.5$ (middle left two), 
       FENE-CR of $L=$5 and $\lambda_1=1$ (middle right two), and  Oldroyd-B and $\lambda_1=1$ (right two).}    \label{fig2d}
\end{figure}     
\begin{figure}
\begin{center} 
\includegraphics[width=0.245\textwidth]{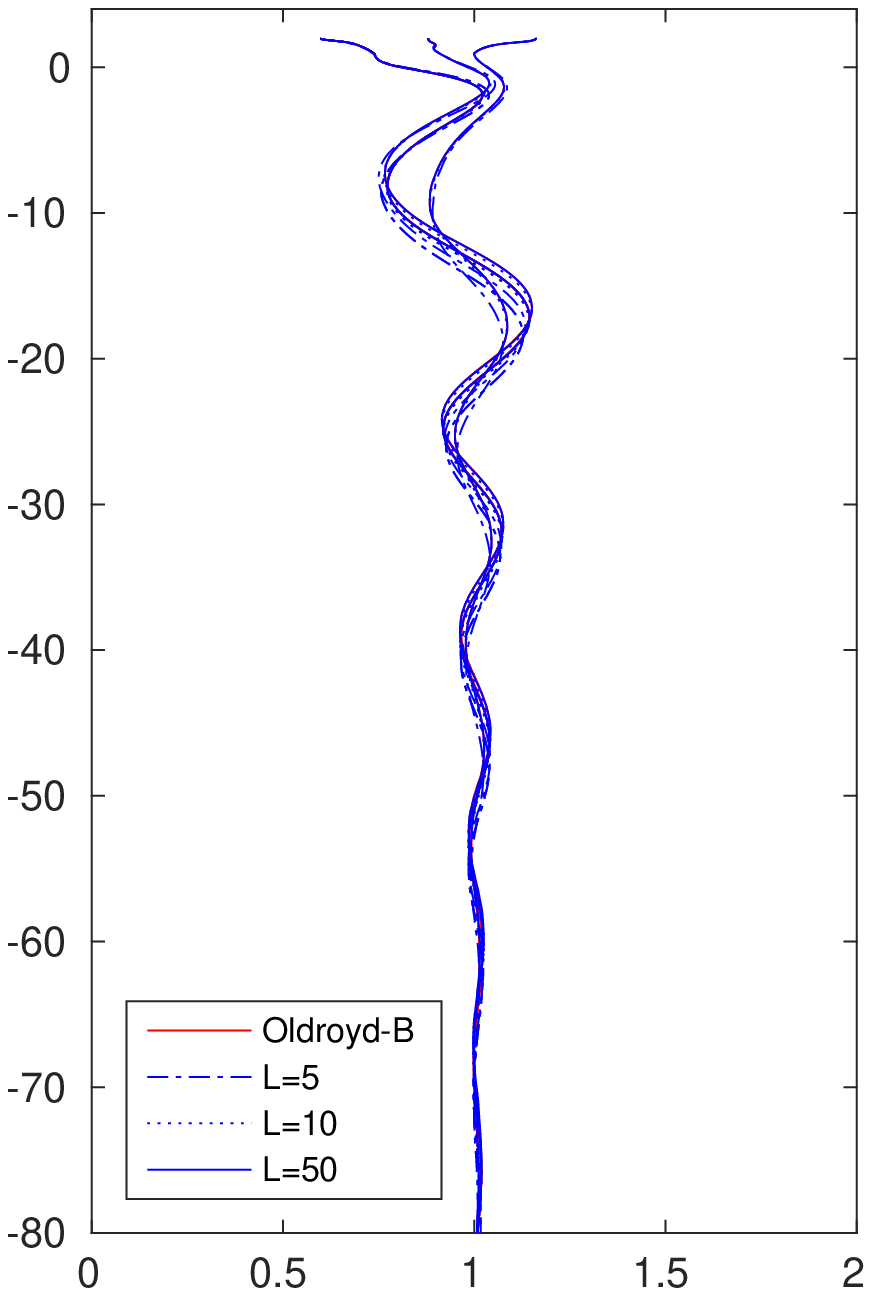}\
\includegraphics[width=0.245\textwidth]{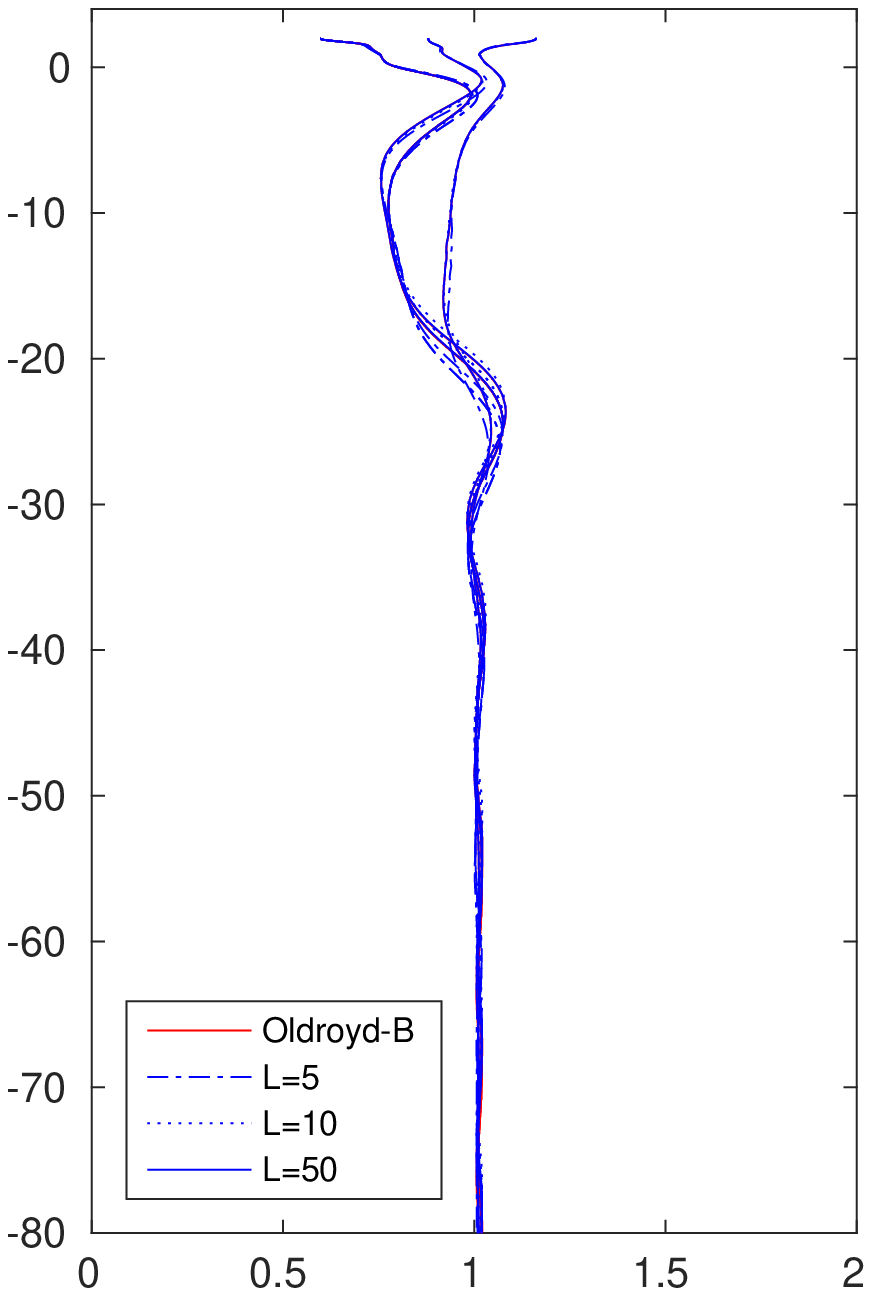}\
\includegraphics[width=0.245\textwidth]{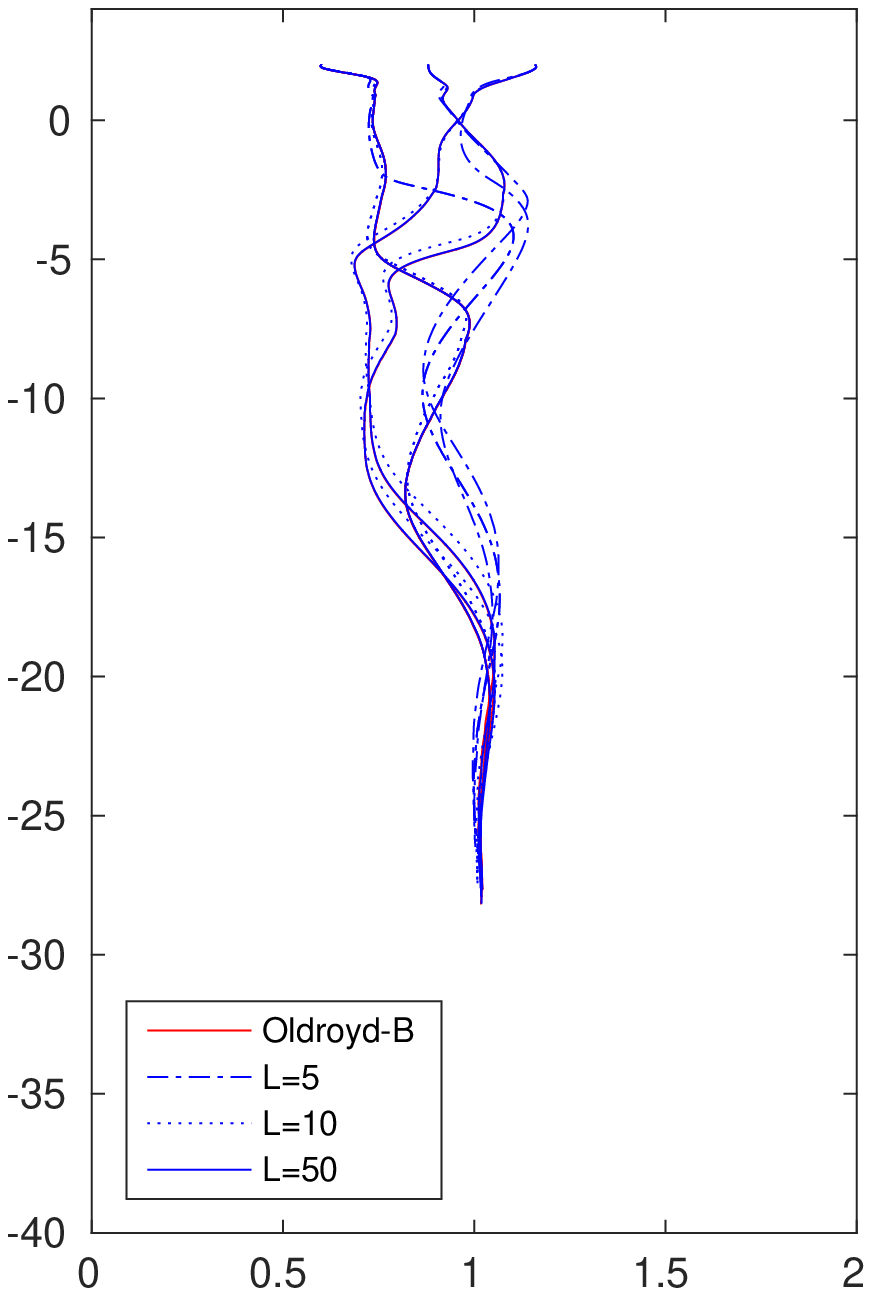}\          
\includegraphics[width=0.245\textwidth]{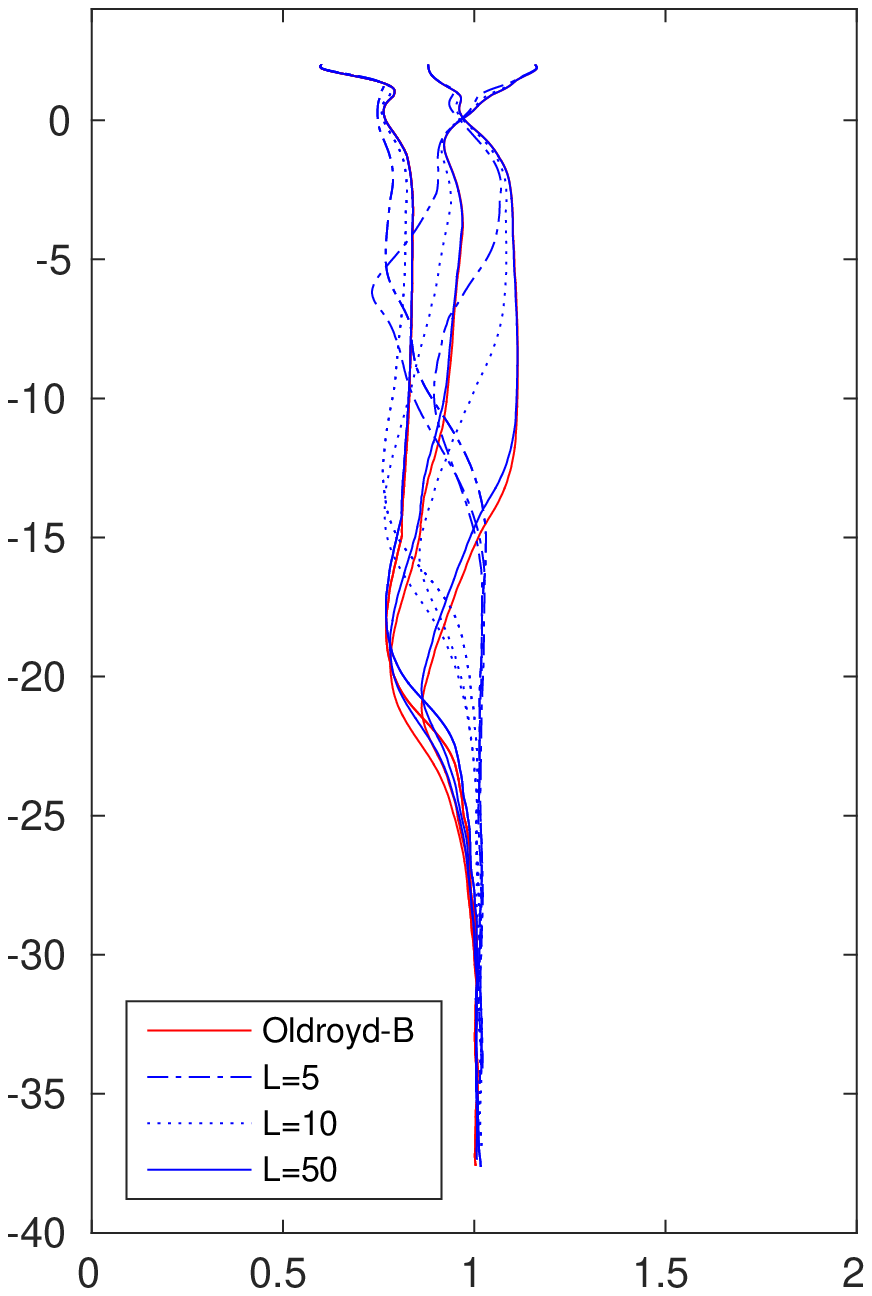}      
\end{center}  
\caption{The disk trajectories: $\rho_s$= 1.0075 and $\lambda_1=0.75$ (left),   $\rho_s$=1.01, $\lambda_1=0.75$ (middle left),
 $\rho_s$=1.0075, $\lambda_1=1.5$ (middle right),  and    $\rho_s$=1.01 $\lambda_1=1.5$ (right).}  \label{fig3a}
\end{figure}

\begin{figure}[!tp]
\begin{center} 
\includegraphics[width=0.119\textwidth]{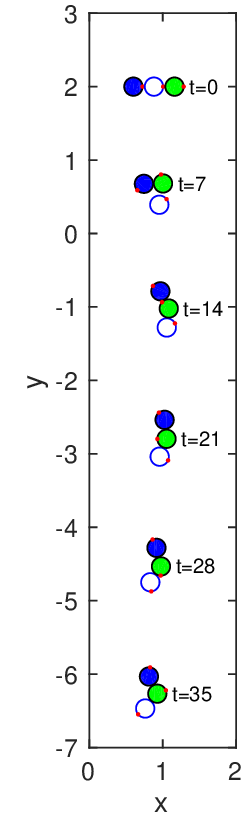}
\includegraphics[width=0.119\textwidth]{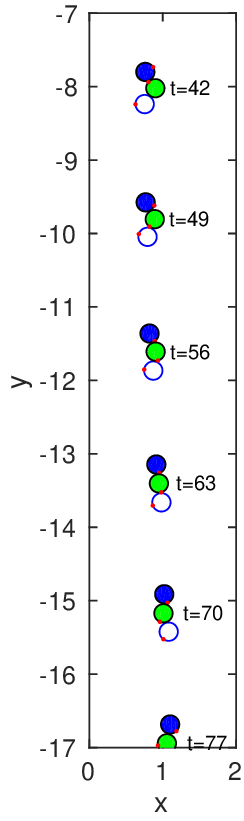}
\includegraphics[width=0.119\textwidth]{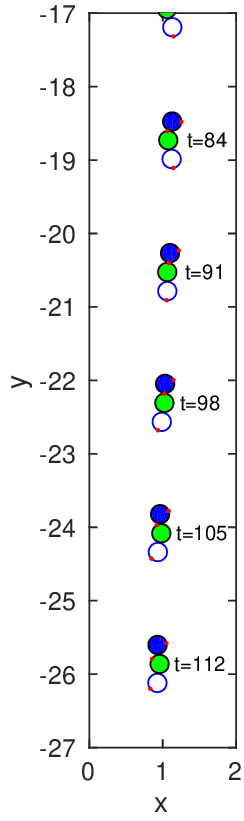}
\includegraphics[width=0.119\textwidth]{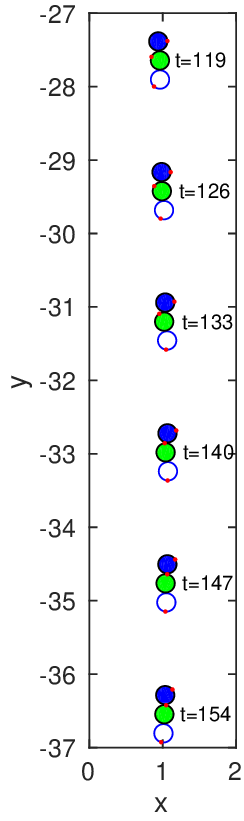}
\includegraphics[width=0.119\textwidth]{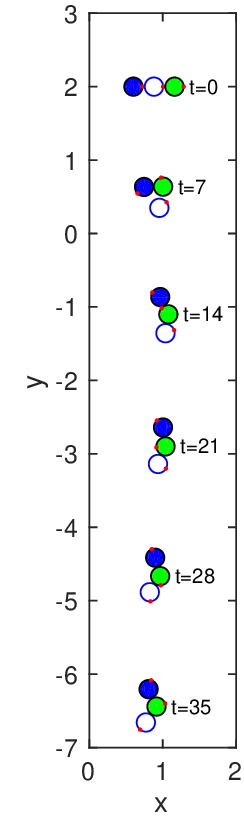}
\includegraphics[width=0.119\textwidth]{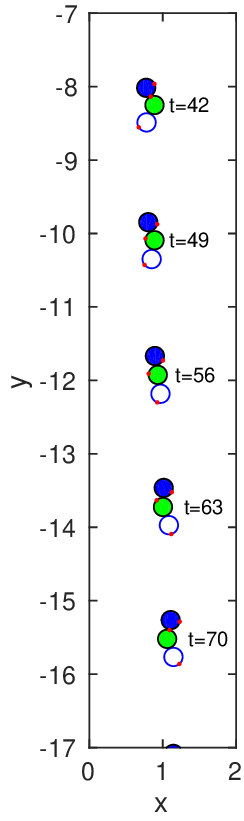}     
\includegraphics[width=0.119\textwidth]{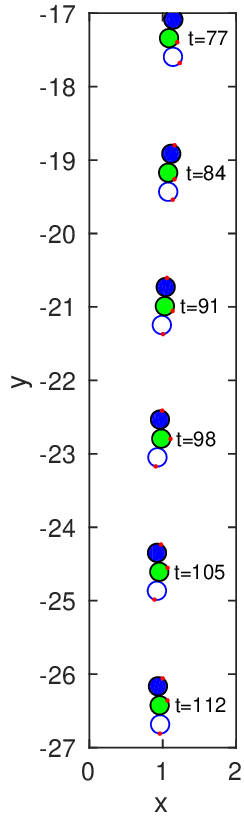}         
\includegraphics[width=0.119\textwidth]{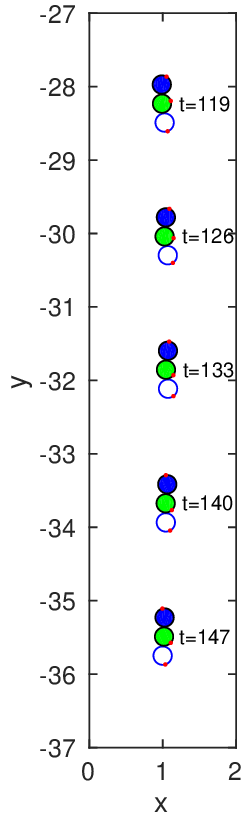}         
\end{center}  
\caption{Positions of three disks  for $\rho_s$= 1.0075: FENE-CR of $L=$5 \& $\lambda_1=0.75$, Re=0.244, M=0.431, De=0.762,  and  E=3.12  (left four) 
and  Oldroyd-B \& $\lambda_1=0.75$, Re=0.259, M=0.440, De=0.777, E=3.12 (right four).}    \label{fig3b}
% \end{figure}                                
% \begin{figure}
\begin{center} 
\includegraphics[width=0.13\textwidth]{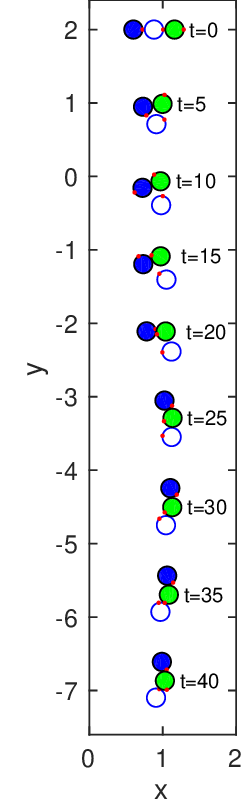}
\includegraphics[width=0.13\textwidth]{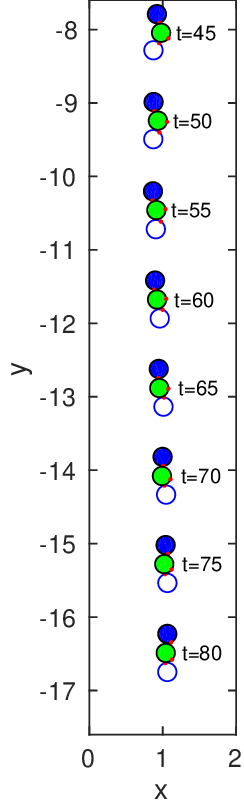}
\includegraphics[width=0.13\textwidth]{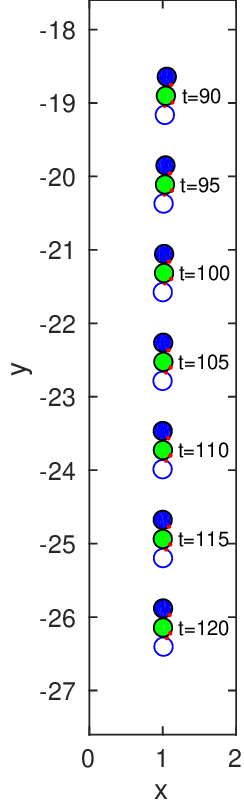}
\includegraphics[width=0.13\textwidth]{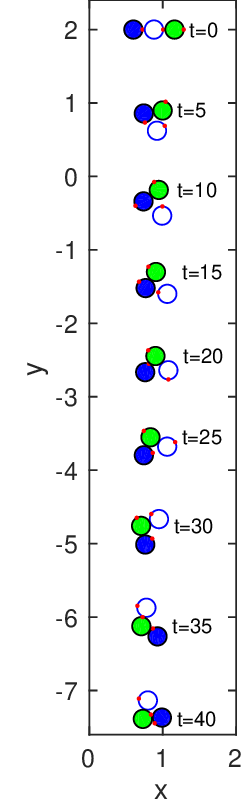}
\includegraphics[width=0.13\textwidth]{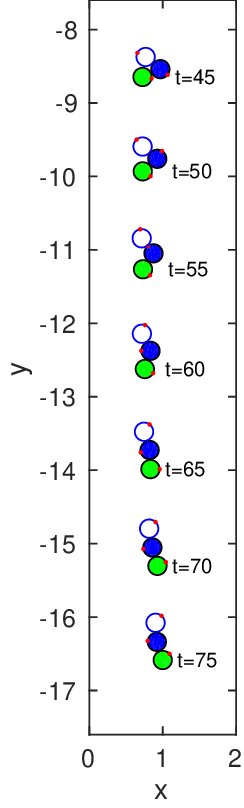}     
\includegraphics[width=0.13\textwidth]{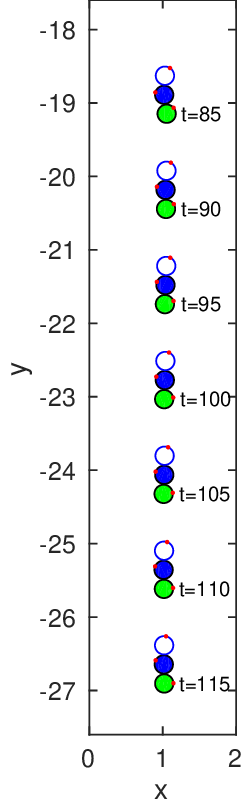}         
\end{center}  
\caption{Positions of three disks  for $\rho_s$= 1.0075:  FENE-CR of $L=$5 \& $\lambda_1=1.5$, Re=0.232, M=0.579, De=1.447, E=6.24  (left three)
and   Oldroyd-B \& $\lambda_1=1.5$, Re=0.248, M=0.620, De=1.550, E=6.24 (right three).}    \label{fig3c}
\end{figure}

For the cases of two disks sedimenting in a vertical channel of  infinite length filled with a viscoelastic fluid, 
the computational domain is $\Omega= (0,1)\times(0,6)$ initially and then it moves vertically with the mass center of the 
lower disk between two disks. The two disk diameters are $d=$0.25 
and the initial position of the disk centers are at (0.35, 2.5) and (0.65, 2.5), respectively. The disk density 
$\rho_s$ is 1.01 and the fluid density $\rho_f$ is 1. The fluid viscosity 
$\eta_1$ is 0.2.  The relaxation time $\lambda_1$ is either 0.5 or 1 and the retardation time $\lambda_2$ is $\lambda_1/4$.
Hence the elasticity number E are 1.6 and 3.2 for $\lambda_1=0.5$ and 1, respectively. The maximal polymer extension $L$ is  
either 5, 10, or 50 for the  FENE-CR model. The mesh sizes for the velocity field, conformation tensor and pressure are 
$h=1/96$, 1/96, and 1/48, respectively; and  the time step is 0.0004. Fig. \ref{fig2a} shows that  the trajectories of two 
disks settling in either Oldroyd-B or FENE-CR fluids are almost identical  for $\lambda_1=0.5$  and $L$=5, 10, and 50.
Similar behaviors are also observed for the vertical and horizontal particle translation velocities shown in Figs.  \ref{fig2a}.
But,    in Fig.  \ref{fig2b}, the two disk trajectories for  $\lambda_1=1$ and $L$=5 and 10 are slightly different from those for $\lambda_1=1$ 
and $L=50$ and those for Oldroyd-B fluid. Similar behaviors for the cases of  $L$=5 and 10 are also observed for the vertical and horizontal 
particle translation velocities shown in Fig.  \ref{fig2b}.  For all cases, two disks attract to each other first, 
form a horizontal chain and then  its broadside is turn into the falling direction; it is different from the well known phenomenon 
called drafting, kissing and tumbling for disks settling in Newtonian fluid \cite{fortes1987}. In the position snapshots shown in 
Fig. \ref{fig2d},  for the case of $L=5$ and E=1.6 two disks move slowest due to the smaller relaxation time and shorter extension 
value of $L$ and, on the other hand, for the case of Oldroyd-B  fluid and E=3.2, two disks move further down in the channel. But for 
the both cases of $L=50$, the particle motions are almost identical to those in Oldroyd-B fluid as 
in  Figs. \ref{fig2a} and  \ref{fig2b}.

For the cases of three disks sedimenting in a vertical channel of  infinite length filled with a viscoelastic fluid, 
the computational domain is $\Omega= (0,2)\times(0,5)$ initially and then it moves vertically with the mass center of the 
lowest disk among three disks. The three disk diameters are $d=$0.25 and the initial disk centers are 
(0.6, 2), (0.88, 2) and (1.16, 2), respectively. The disk density $\rho_s$ is either 1.0075 or 1.01 and the fluid density $\rho_f$ is 1. 
The fluid viscosity $\eta_1$ is 0.26.  The relaxation time $\lambda_1$ is either 0.75 or 1.5 and the retardation time $\lambda_2$ is $\lambda_1/8$. 
Hence the elasticity number E are 3.12 and 6.24 for $\lambda_1=0.75$ and 1.5, respectively. The maximal polymer extension $L$ is either 
5, 10, or 50 for the  FENE-CR model. The mesh sizes for the velocity field,  conformation tensor and pressure are $h=1/96$, 1/96, and 1/48, 
respectively; and the time step is 0.0004. Fig. \ref{fig3a} shows that the trajectories of three disks of both densities settling in 
either Oldroyd-B or FENE-CR fluids are close to each other  for E=3.12, $L$=10 and 50. But at E=6.24, the three disk trajectories for $L$=5
are very different from those for $L=50$ and those for Oldroyd-B fluid.  For all cases shown in Figs. \ref{fig3b} and \ref{fig3c},  
the middle disk moves downward faster first and the other two  are drafted toward this leading disk. Then they rearrange themselves and  
form a curved chain. Finally the curved chain straighten out due to the large normal stress next to the middle disk in the curved chain
(as discussed in \cite{huang1998}). For the cases of $L=50$, the particle trajectories  are almost identical to those of Oldroyd-B fluid 
as in  Fig. \ref{fig3a}.  

The above numerical results suggest that the effect of the shorter extensibility in the FENE-CR model on the particle
settling trajectory can be enhanced by increasing the value of relaxation time $\lambda_1$. It is known that when the elasticity 
number E=De/Re is larger than the critical value ($O(1)$) and the Mach number M=$\sqrt{{\rm De Re}}$ is less than the critical value, 
the long particle settling in an Oldroyd-B fluid can turn its broadside parallel to the falling direction (\cite{huang1998}, \cite{Liu1993}).  
For two disk and three disk cases 
considered in this section, the values of the elasticity number are  large than the critical values and those of the Mach number 
are smaller than 1  and  particle chain is always formed with its ``broadside'' parallel parallel to the flow direction. Thus, at least for the cases 
and the values  of $L$ considered in this section, we have obtained that the polymer extension limit $L$ has no effect on a short vertical chain 
formation of particles; but the intermediate dynamics of particle interaction before having a vertical chain can be different 
for smaller values of $L$ when increasing the relaxation time $\lambda_1$.

\begin{figure} [tp]
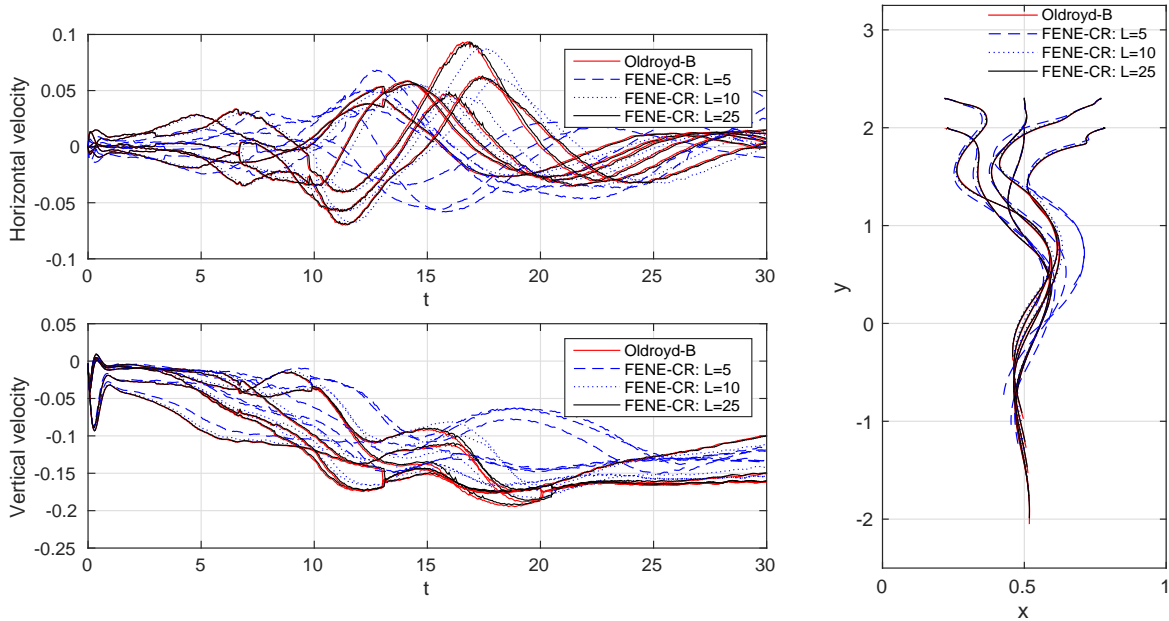

\begin{minipage}{4.2in}
%         \begin{figure}
\includegraphics[width=4.0in]{6disk-case3E5d40-xcrv-comp.eps}\\
\includegraphics[width=4.0in]{6disk-case3E5d40-ycrv-comp.eps}\
%         \end{figure}
  \end{minipage}      
  \begin{minipage}{2in}
%         \begin{figure}
\includegraphics[width=1.8in]{6disk-case3E5d40-position-comp.eps}  
%         \end{figure}
  \end{minipage}       
\caption{Histories of the particle velocity (left) and trajectories of six disks (right) for $\lambda_1$=1.3.}  \label{fig4d}
\end{figure}

\begin{figure} [tp]
\begin{center}
\leavevmode
\epsfxsize=0.48in
\epsffile{6disk-case3E5d40-OL-t=02.eps}  
\epsfxsize=0.48in                   
\epsffile{6disk-case3E5d40-OL-t=10.eps}  
\epsfxsize=0.48in                   
\epsffile{6disk-case3E5d40-OL-t=12.eps}  
\epsfxsize=0.48in                   
\epsffile{6disk-case3E5d40-OL-t=14.eps}  
\epsfxsize=0.48in                   
\epsffile{6disk-case3E5d40-OL-t=16.eps}  
\epsfxsize=0.48in                   
\epsffile{6disk-case3E5d40-OL-t=18.eps}  
\epsfxsize=0.48in                   
\epsffile{6disk-case3E5d40-OL-t=20.eps}  
\epsfxsize=0.48in                   
\epsffile{6disk-case3E5d40-OL-t=22.eps}  
\epsfxsize=0.48in                   
\epsffile{6disk-case3E5d40-OL-t=24.eps}  
\epsfxsize=0.48in                   
\epsffile{6disk-case3E5d40-OL-t=26.eps}  
\epsfxsize=0.48in                   
\epsffile{6disk-case3E5d40-OL-t=28.eps}  
\epsfxsize=0.48in
\epsffile{6disk-case3E5d40-OL-t=30.eps} \\
\epsfxsize=0.48in
\epsffile{6disk-case3E5d40-L005-t=02.eps}  
\epsfxsize=0.48in                     
\epsffile{6disk-case3E5d40-L005-t=10.eps}  
\epsfxsize=0.48in                     
\epsffile{6disk-case3E5d40-L005-t=12.eps}  
\epsfxsize=0.48in                     
\epsffile{6disk-case3E5d40-L005-t=14.eps}  
\epsfxsize=0.48in                     
\epsffile{6disk-case3E5d40-L005-t=16.eps}  
\epsfxsize=0.48in                     
\epsffile{6disk-case3E5d40-L005-t=18.eps}  
\epsfxsize=0.48in                     
\epsffile{6disk-case3E5d40-L005-t=20.eps}  
\epsfxsize=0.48in                     
\epsffile{6disk-case3E5d40-L005-t=22.eps}  
\epsfxsize=0.48in                     
\epsffile{6disk-case3E5d40-L005-t=24.eps}  
\epsfxsize=0.48in                     
\epsffile{6disk-case3E5d40-L005-t=26.eps}  
\epsfxsize=0.48in                     
\epsffile{6disk-case3E5d40-L005-t=28.eps}  
\epsfxsize=0.48in
\epsffile{6disk-case3E5d40-L005-t=30.eps} 
\end{center} 
\kern -2ex
\caption{Snapshots  of the positions of six disks at $t=2$, 10, 12, 14, 16, 18, 20, ,22, 24,  26, 28, and 30 for $\lambda_1$=1.3  in an  
Oldroyd-B fluid (top)  and  a FENE-CR fluid with $L=5$ (bottom).} \label{fig4a}
% \end{figure}
% \begin{figure}[th]
\begin{center}
\leavevmode
\epsfxsize=0.48in
\epsffile{6disk-case3E6d24-OL-t=02.eps}  
\epsfxsize=0.48in                   
\epsffile{6disk-case3E6d24-OL-t=10.eps}  
\epsfxsize=0.48in                   
\epsffile{6disk-case3E6d24-OL-t=12.eps}  
\epsfxsize=0.48in                   
\epsffile{6disk-case3E6d24-OL-t=14.eps}  
\epsfxsize=0.48in                   
\epsffile{6disk-case3E6d24-OL-t=16.eps}  
\epsfxsize=0.48in                   
\epsffile{6disk-case3E6d24-OL-t=18.eps}  
\epsfxsize=0.48in                   
\epsffile{6disk-case3E6d24-OL-t=20.eps}  
\epsfxsize=0.48in                   
\epsffile{6disk-case3E6d24-OL-t=22.eps}  
\epsfxsize=0.48in                   
\epsffile{6disk-case3E6d24-OL-t=24.eps}  
\epsfxsize=0.48in                   
\epsffile{6disk-case3E6d24-OL-t=26.eps}  
\epsfxsize=0.48in                   
\epsffile{6disk-case3E6d24-OL-t=28.eps}  
\epsfxsize=0.48in
\epsffile{6disk-case3E6d24-OL-t=30.eps} \\
\epsfxsize=0.48in
\epsffile{6disk-case3E6d24-L005-t=02.eps}  
\epsfxsize=0.48in                     
\epsffile{6disk-case3E6d24-L005-t=10.eps}  
\epsfxsize=0.48in                     
\epsffile{6disk-case3E6d24-L005-t=12.eps}  
\epsfxsize=0.48in                     
\epsffile{6disk-case3E6d24-L005-t=14.eps}  
\epsfxsize=0.48in                     
\epsffile{6disk-case3E6d24-L005-t=16.eps}  
\epsfxsize=0.48in                     
\epsffile{6disk-case3E6d24-L005-t=18.eps}  
\epsfxsize=0.48in                     
\epsffile{6disk-case3E6d24-L005-t=20.eps}  
\epsfxsize=0.48in                     
\epsffile{6disk-case3E6d24-L005-t=22.eps}  
\epsfxsize=0.48in                     
\epsffile{6disk-case3E6d24-L005-t=24.eps}  
\epsfxsize=0.48in                     
\epsffile{6disk-case3E6d24-L005-t=26.eps}  
\epsfxsize=0.48in                     
\epsffile{6disk-case3E6d24-L005-t=28.eps}  
\epsfxsize=0.48in
\epsffile{6disk-case3E6d24-L005-t=30.eps} 
\end{center} 
\kern -2.1ex
\caption{Snapshots  of the positions of six disks at $t=2$, 10, 12, 14, 16, 18, 20,  22, 24,  26, 28, and 30 for $\lambda_1$=1.5  in an  
Oldroyd-B fluid (top)  and  a FENE-CR fluid with $L=5$  (bottom) and 
the associated numbers are Re=0.149, M=0.3721, De=0.9295, E=6.24 and Re=0.1278, M=0.3194, De=0.7978, E=6.24 for Oldroyd-B fluid and FENE-CR 
of L=5, respectively.} \label{fig4c}
\end{figure}

\subsection{Several settling disks}

\begin{figure}  [!tp]
\begin{center}
\leavevmode
\epsfxsize=0.45in
\epsffile{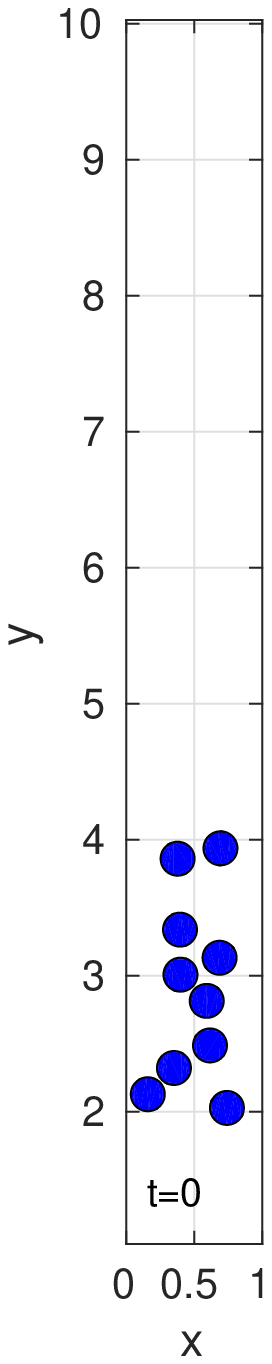}  
\epsfxsize=0.45in                     
\epsffile{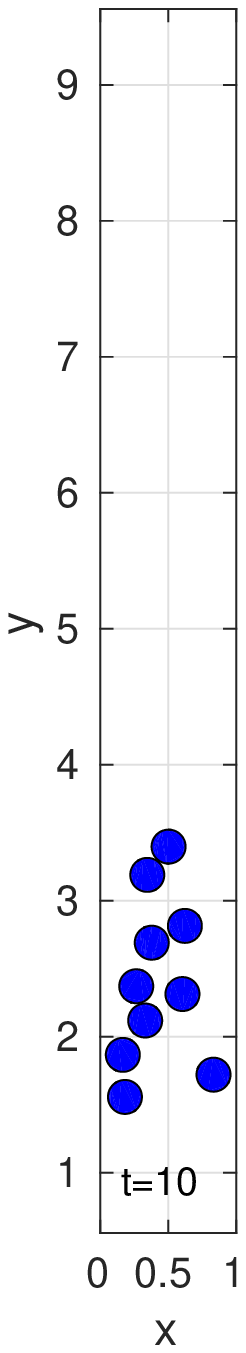}  
\epsfxsize=0.45in                     
\epsffile{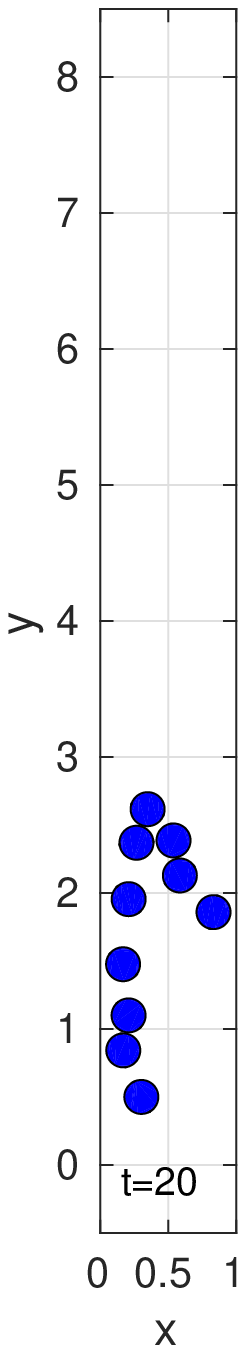}  
\epsfxsize=0.45in                     
\epsffile{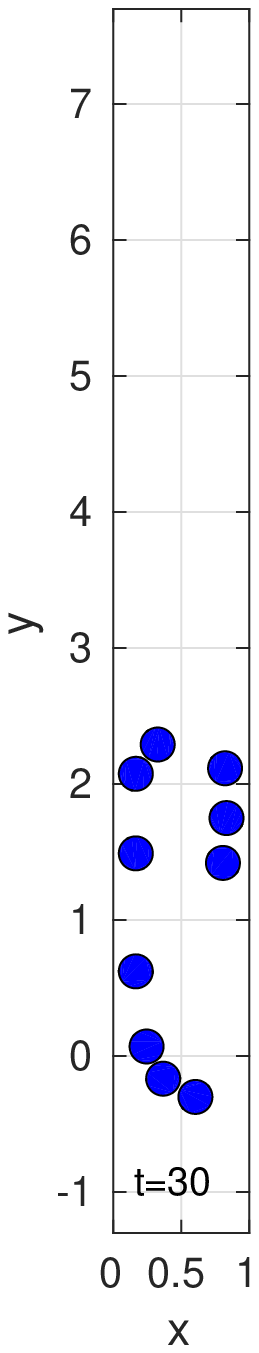}  
\epsfxsize=0.45in                     
\epsffile{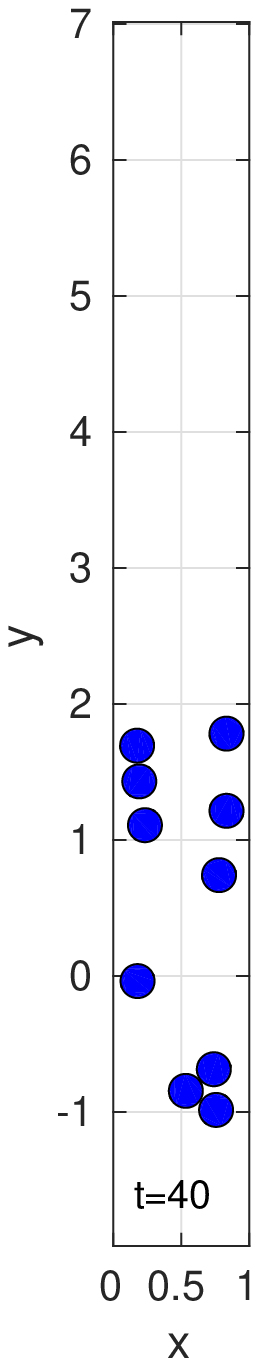}  
\epsfxsize=0.45in                     
\epsffile{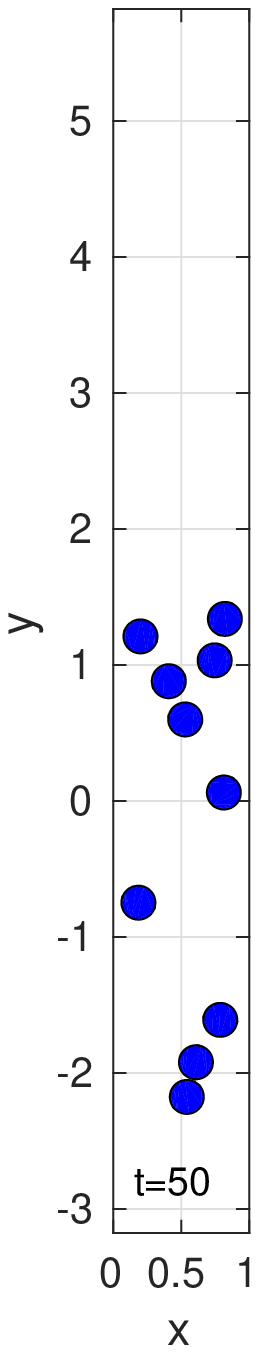}  
\epsfxsize=0.45in                     
\epsffile{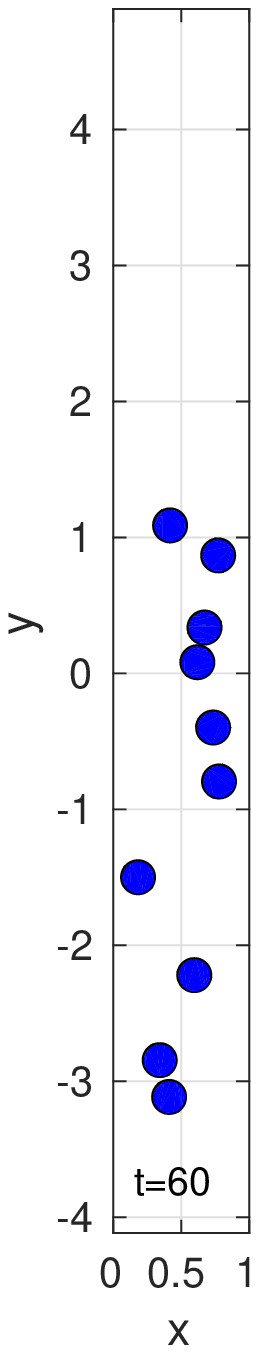}  
\epsfxsize=0.45in                     
\epsffile{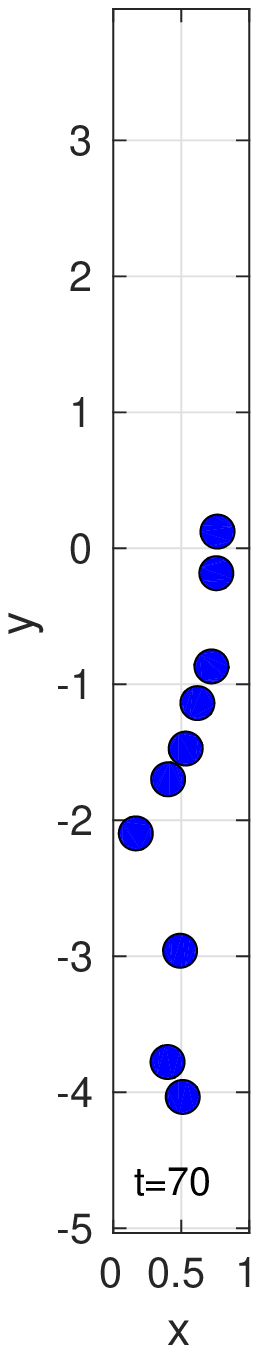}  
\epsfxsize=0.45in                     
\epsffile{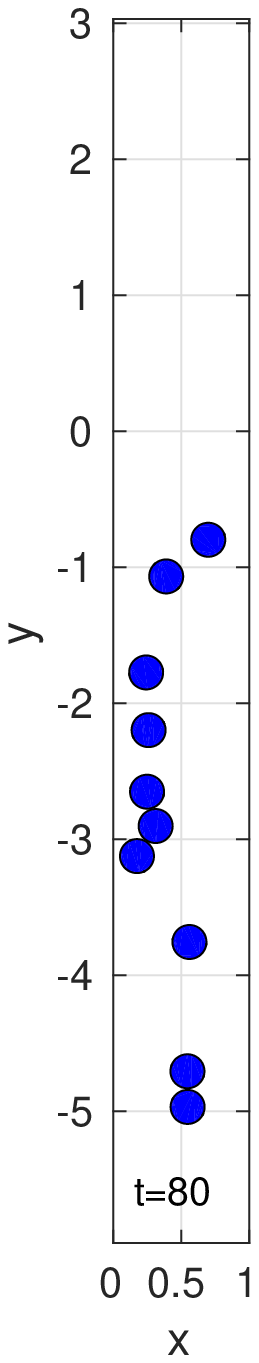} \\
\epsfxsize=0.45in
\epsffile{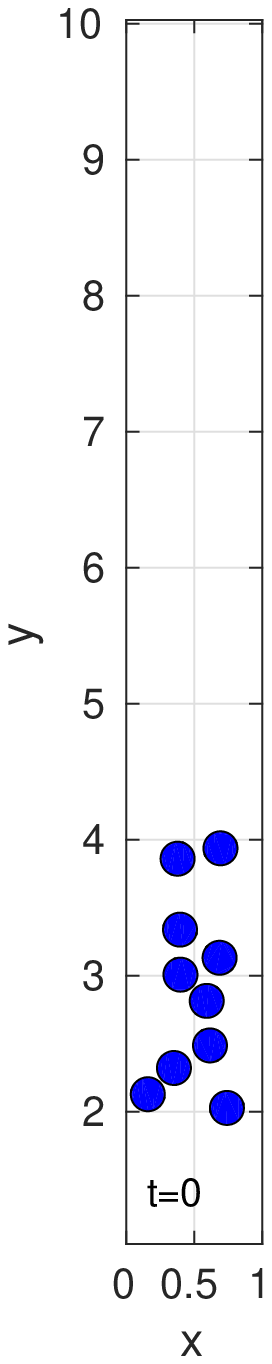}  
\epsfxsize=0.45in                     
\epsffile{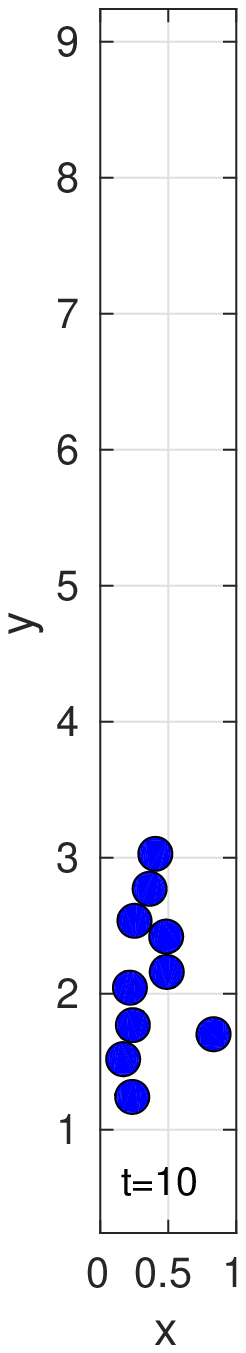}  
\epsfxsize=0.45in                     
\epsffile{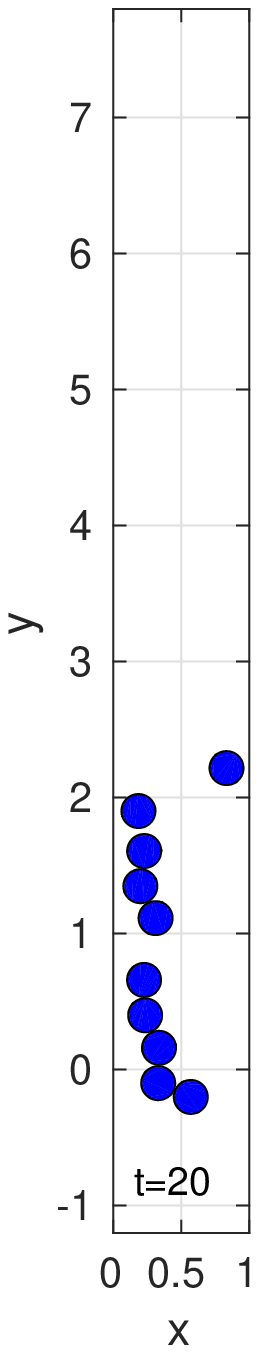}  
\epsfxsize=0.45in                     
\epsffile{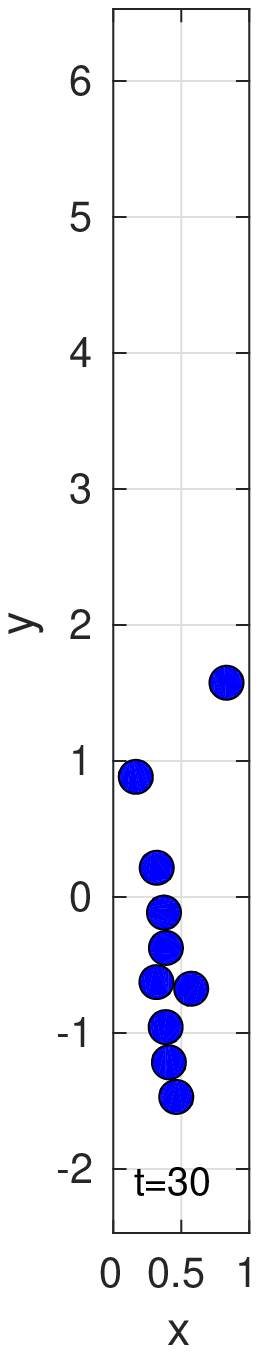}  
\epsfxsize=0.45in                     
\epsffile{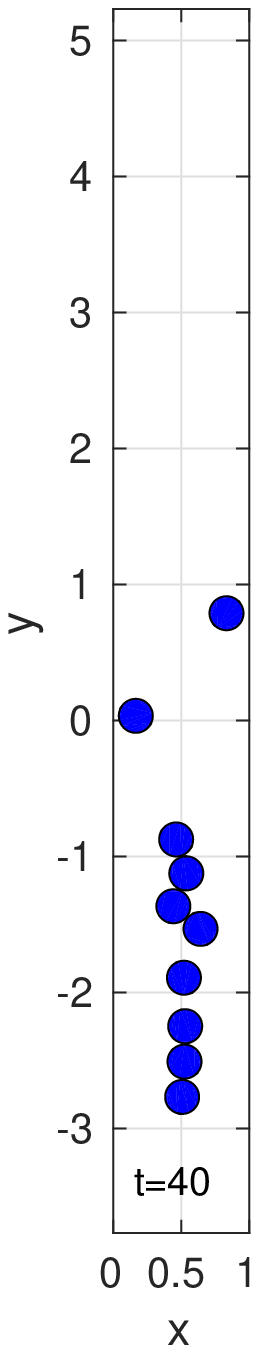}  
\epsfxsize=0.45in                     
\epsffile{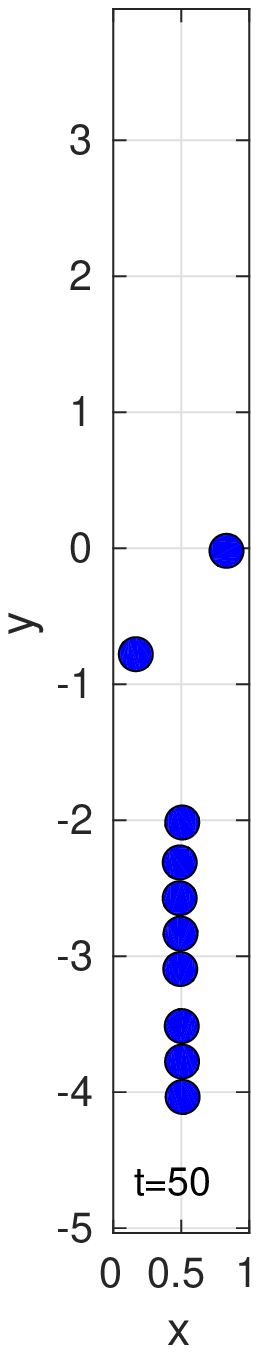}  
\epsfxsize=0.45in                     
\epsffile{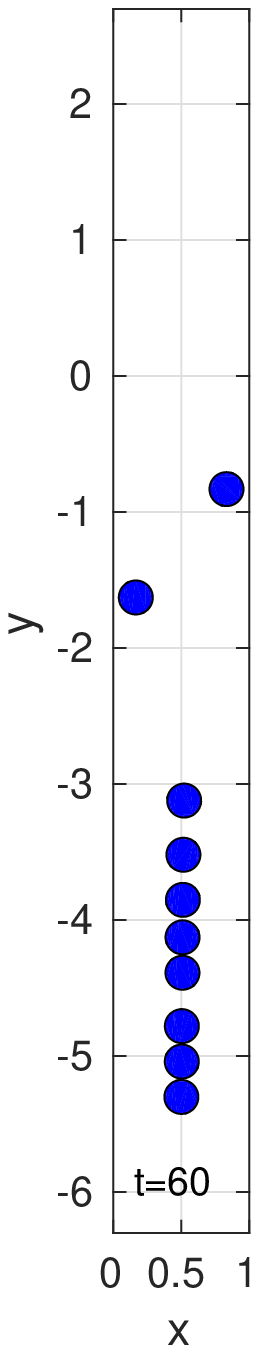}  
\epsfxsize=0.45in                     
\epsffile{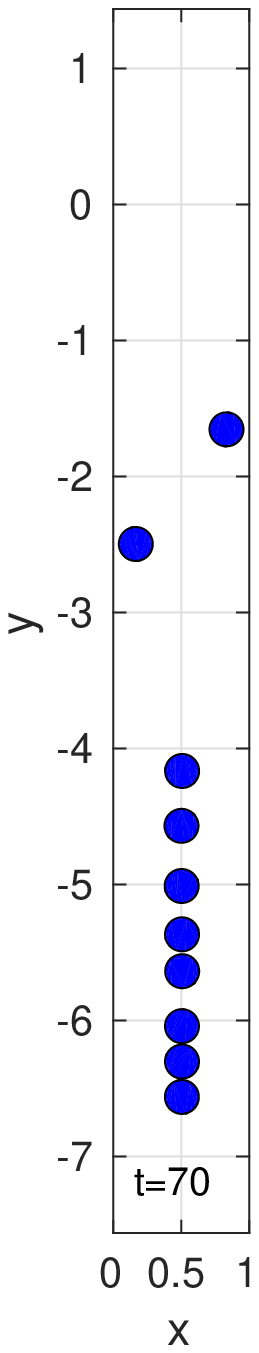}  
\epsfxsize=0.45in                     
\epsffile{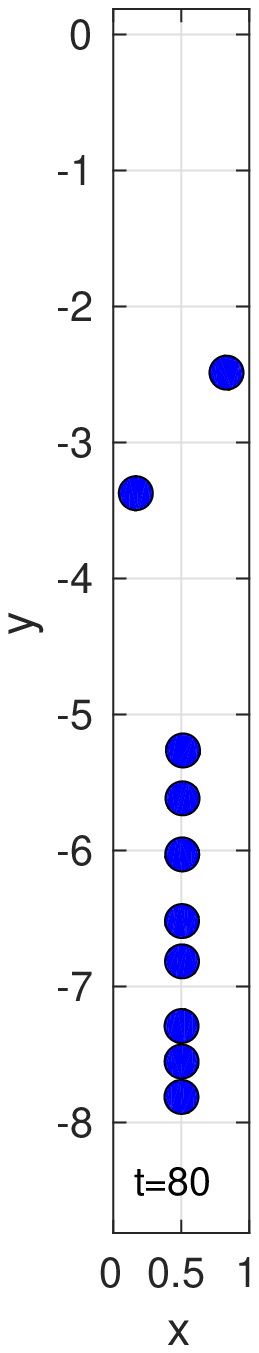} \\
\epsfxsize=0.45in
\epsffile{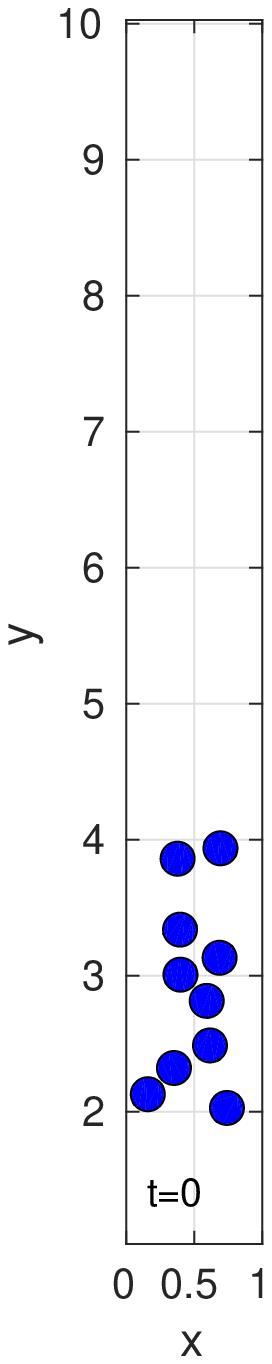}  
\epsfxsize=0.45in                   
\epsffile{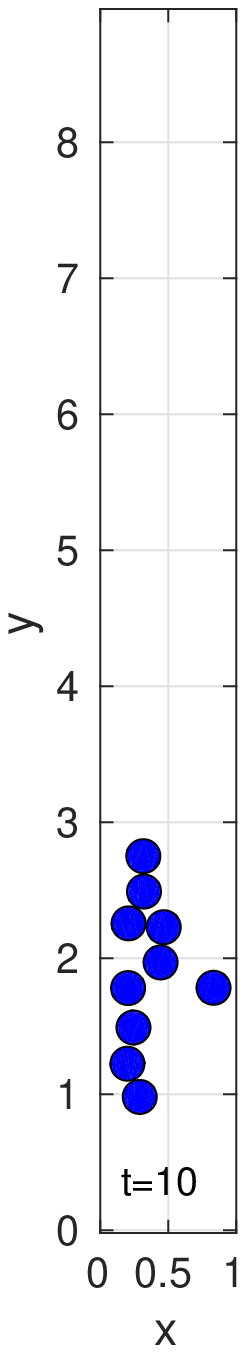}  
\epsfxsize=0.45in                   
\epsffile{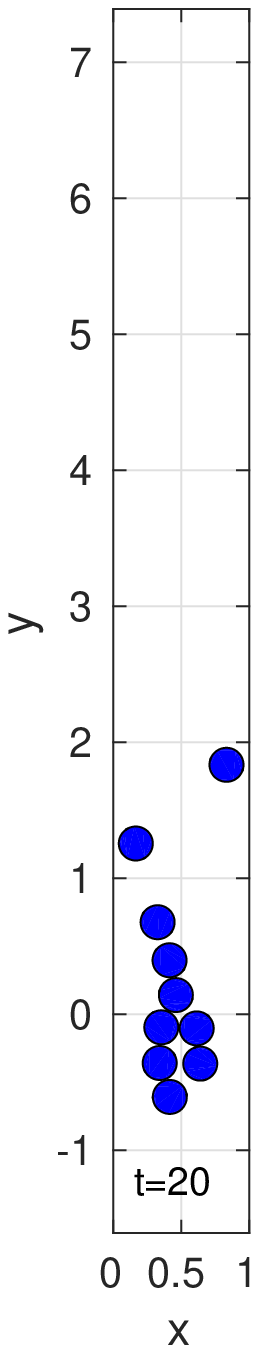}  
\epsfxsize=0.45in                   
\epsffile{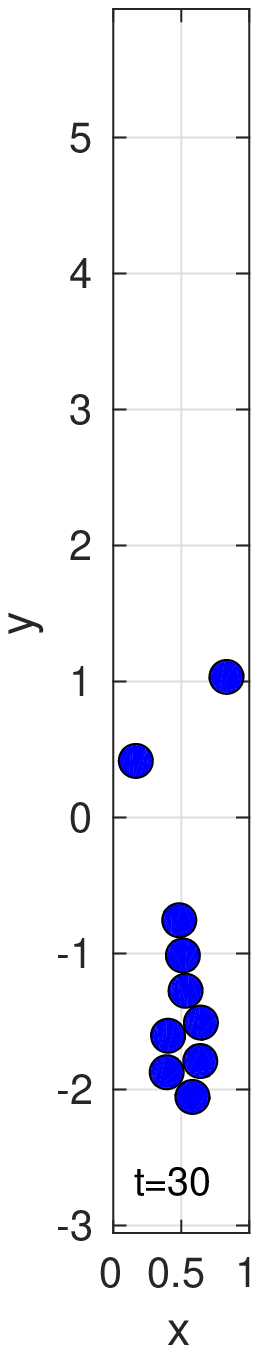}  
\epsfxsize=0.45in                   
\epsffile{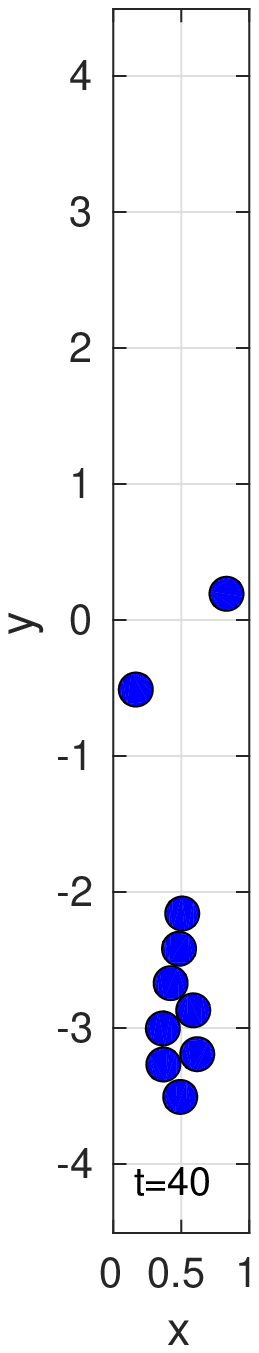}  
\epsfxsize=0.45in                   
\epsffile{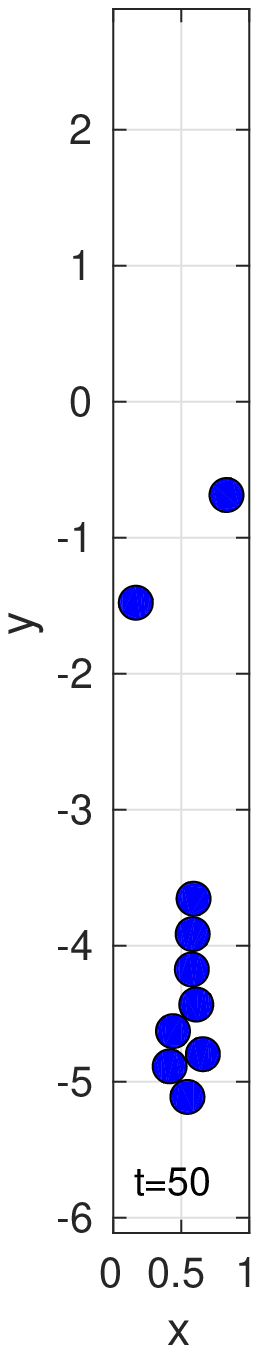}  
\epsfxsize=0.45in                   
\epsffile{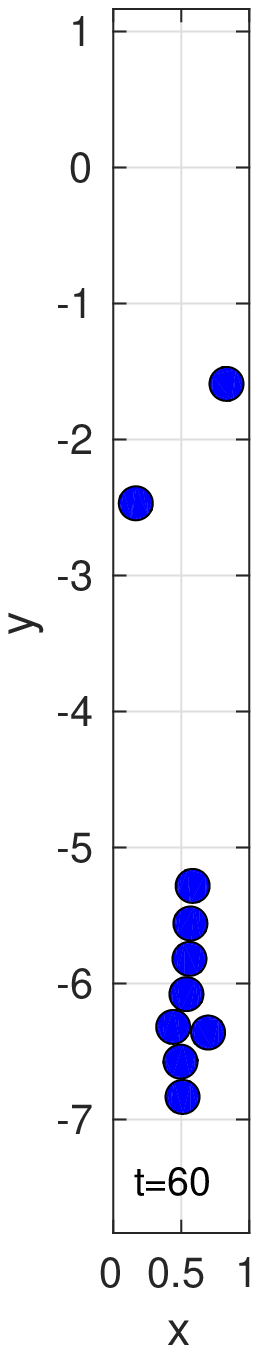}  
\epsfxsize=0.45in                   
\epsffile{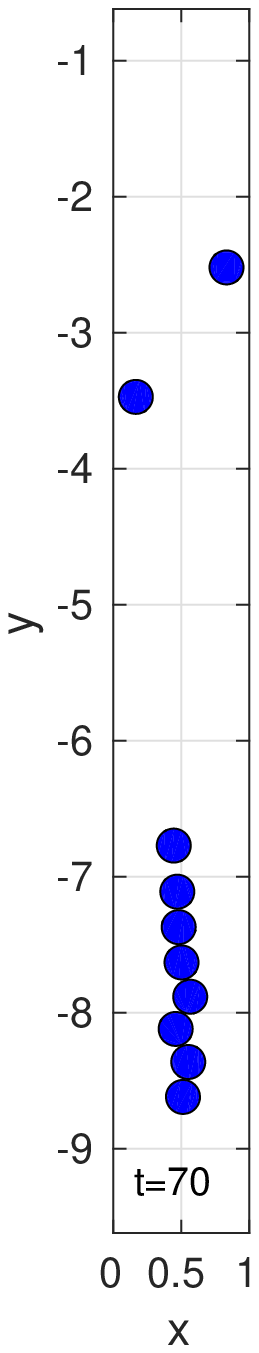}  
\epsfxsize=0.45in                   
\epsffile{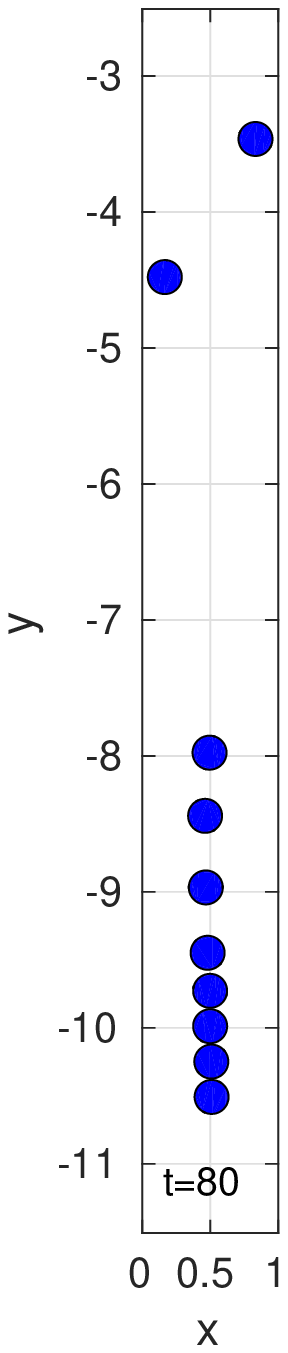} 
\end{center} 
\kern -2ex
\caption{{Snapshots  of the positions of ten disks at $t=0$, 10, 20, 30, 40, 50, 60, 70  and 80  in  FENE-CR fluid for $L=2$  (top) and $L=5$ (middle) and in
Oldroyd-B fluid (bottom).  The associated numbers are Re=0.1539, M=0.4439, De=1.2804, E=8.32 for Oldroyd-B fluid; Re=0.09626,M=0.2777,De=0.8009,
E=8.32 for FENE-CR fluid and L=2, and Re=0.1092, M=0.3150, De=0.9085, E=8.32 for  FENE-CR fluid and L=5.}} \label{fig4e}
\end{figure}

To find out the effect of the polymer extension limit $L$ on the formation of a long chain of particles, 
the first case in this section concerns six disks of diameter $d=$0.25 sedimenting in a 
channel filled with either an Oldroyd-B fluid or a FENE-CR fluid of viscosity $\eta_1=$0.26. 
The channel is infinitely long and has a width of 1. The computational domain is $\Omega=(0,1)\times (0,7)$ 
initially and then moves down with the  mass center of the lowest of the six particles.  
The initial positions of the disks are (0.23, 2.0), (0.5,2.0), (0.78, 2.0), (0.22, 2.30), (0.5, 2.3), and (0.77, 2.3). 
The disk density is $\rho_s$=1.01 and the fluid density is $\rho_f$=1. The relaxation time and retardation time 
are $\lambda_1$=1.3 and $\lambda_2=\lambda_1/8$, respectively.  The mesh sizes for the velocity field, conformation
tensor and pressure are $h=1/96$, 1/96, and 1/48, respectively; and the time step is 0.0004.
In our simulations, all six particles in an Oldroyd-B fluid are lined up along the flow direction, 
agreeing thus the known observations and experiments. Fig. \ref {fig4a} gives the snapshots  at various 
moments of time  of the particles lining up phenomenon. We can see that, after drafting, kissing and chaining, 
the six particles form approximately a straight line at $t = 20$. Then a chain of 5 disks is maintained from $t=22$ to 28;
and at the same time duration the trailing particle has been separated from the leading five particles. This observation 
agrees with experiments showing that, sometimes, the last particle in the chain gets detached as  discussed 
in  \cite{patankar2000}. It is known that a long chain falls faster than a single particle in the fluid. This long body effect
tends to detach the last particle from the chain.  The average terminal velocity is 0.1535 for $26\le t \le 30$ , 
the Reynolds number is Re=0.1476, the Deborah number is De=0.7981, the elasticity number is E=5.408 
and the Mach number is M=0.3432. For the FENE-CR model for the polymer extension limit $L=5$, since the 
viscoelastic fluid has a shorter polymer extension limit, it can not hold all six disks together as shown in Fig.  \ref {fig4a} 
for $t \ge 16$; but instead two chains of three disks are formed and maintained.   For this case, the average 
terminal velocity is 0.1317 for $26\le t \le 30$, and the associated numbers are  Re=0.1266,   De=0.6847,   
E=5.408, and   M=0.2944. The particle velocities and trajectories for different values of $L$ and  $\lambda_1=1.3$ 
in Fig. \ref{fig4d} show that,  as $L$ is about 25, the dynamics of six disks in FENE-CR fluid is almost identical to the 
one in Oldroyd-B fluid. For slightly larger $\lambda_1=1.5$, the particle dynamics and chain formation are similar to 
those for  $\lambda_1=1.3$; however those chains of disks in Fig. \ref{fig4c}  are straighten out faster 
comparing to those in  Fig. \ref{fig4a}  by stronger normal stress due to the larger value of E.

{ In the second case of this section, we  have increased the number of disks to ten and have kept all other parameters the same
except that the computational domain is $\Omega=(0,1)\times (0,16)$, the relaxation time is  $\lambda_1$=2 (E=8.32), and the  ten disk mass centers  
are randomly chosen  in the region $(0,1)\times (2,4)$ initially.  This  initial position give us some  computational results 
concerning the effect of polymer extension limit $L$ on the agglomeration and chain of particles. Fig. \ref{fig4e} 
is obtained for Oldroyd-B and FENE-CR viscoelastic fluids, respectively. In Oldroyd-B fluid, the positions of 10 disks 
at different instants of time show that the agglomeration of particles can be held initially in a  cluster of 8 disks 
for $20 \le t \le 60$; but  after the cluster becomes a long chain around $t=70$,  the formation of a long chain can not be kept 
due to the detachment of the trailing particle as discussed in the previous case of six disks.  The results of the ten particles settling in an FENE-CR for $L=2$ 
are quite different from those  in Oldroyd-B fluid. All disks spread out most of time. The results of the ten particles settling in an FENE-CR for $L=5$ are 
different from the those in Oldroyd-B fluid since these particles are still relatively easier to break away from the chains and clusters.  There are two clusters for 
$20 \le t \le 40$; once the disks in these two clusters all line up, the last disk  in the chain of more than 3 disks  keep breaking away. The numerical results of 
three cases suggest that for smaller values of $L$, the length of the vertical chain is shorter and the size of cluster is smaller.} 

\section{Conclusion}

In this article we present a numerical method for simulating the sedimentation of circular particles
in two-dimensional channel filled with a viscoelastic fluid of FENE-CR type, which is generalized
from a domain/distributed Lagrange multiplier method with a factorization approach for Oldroyd-B fluids 
developed in \cite{Hao2009}.   Numerical results on the vertical chain formation suggest that the polymer extension 
limit $L$ for the FENE-CR fluid has no effect on  the cases of two disks and three disks in two-dimensional narrow 
channel, at least for the values of $L$ considered in this article; but the intermediate dynamics  of particle 
interaction before having a vertical chain can be different for smaller values of $L$ when increasing the relaxation time. 
For six particles sedimenting in FENE-CR type viscoelastic fluid, the formation of disk chains does depend on 
the polymer extension limit $L$. For the smaller values of $L$, two chains of three disks are formed since FENE-CR type 
viscoelastic fluid can not bring them together like the case of these particles settling in a vertical chain formation in Oldroyd-B fluid. 
{Similar results for the case of ten disks are also obtained. The numerical results of several more particle cases suggest 
that for smaller values of $L$, the length of the vertical chain is shorter and the size of cluster is smaller.}  
The next step is to generalize this method to simulate cases of balls sedimenting in a three-dimensional channel filled with either kinds 
of viscoelastic fluid and to study the effect of the elasticity number on the length of particle chain and the size of particle clusters.

\section*{Acknowledgments.} 
 
We acknowledge  the support of NSF (grant DMS-1418308).

\end{document}